\documentclass[prd,twocolumn,showpacs,tightenlines,amssymb,eqsecnum,superscriptaddress,floatfix,byrevtex]{revtex4}
\usepackage{epsfig,latexsym}
\begin{document}

\newcommand{\scri}{{\cal I}}

\title{Retarded radiation from colliding black holes in the close limit}

\author{Sascha Husa}
\affiliation{Albert Einstein Institute, Max Planck Gesellschaft,
             Haus 1, Am M\"uhlenberg, Golm, Germany}
\affiliation{Department of Physics and Astronomy,
             University of Pittsburgh, Pittsburgh, Pennsylvania 15260}

\author{Yosef Zlochower}
\affiliation{Department of Physics and Astronomy,
             University of Pittsburgh, Pittsburgh, Pennsylvania 15260}

\author{Roberto G\'omez}
\affiliation{Department of Physics and Astronomy,
             University of Pittsburgh, Pittsburgh, Pennsylvania 15260}
\affiliation{Pittsburgh Supercomputing Center,
             4400 Fifth Ave., Pittsburgh, Pennsylvania 15213}

\author{Jeffrey Winicour}
\affiliation{Albert Einstein Institute, Max Planck Gesellschaft,
             Haus 1, Am M\"uhlenberg, Golm, Germany}
\affiliation{Department of Physics and Astronomy,
             University of Pittsburgh, Pittsburgh, Pennsylvania 15260}

\begin{abstract}

We use null hypersurface techniques in a new approach to calculate the retarded
waveform from a binary black hole merger in the close approximation. The
process of removing ingoing radiation from the system leads to two notable
features in the shape of the close approximation waveform for a head-on
collision of black holes: (i) an initial quasinormal ringup and (ii) weak
sensitivity to the parameter controlling the collision velocity. Feature (ii)
is unexpected and  has the potential importance of enabling the design of an
efficient template for extracting the gravitational wave signal from the noise.

\end{abstract}

\pacs{04.20.Ex, 04.25.Dm, 04.25.Nx, 04.70.Bw}
\maketitle

\section{Introduction}

We compute the retarded radiation produced by the collision of two black holes
in the close approximation where the collision occurs in the distant past. The
retarded wave corresponds to the perturbation of a Schwarzschild spacetime in
which there is no ingoing radiation. The calculation is carried out by means of
a characteristic evolution algorithm. In previous work \cite{close1} we applied
characteristic evolution to determine the close approximation retarded waveform
from a white hole fission. Here we find that the retarded waveform from a black
hole merger has distinct features which do not appear in the fission waveform:
(i) an initial quasinormal ringup and (ii) weak sensitivity to the parameter
controlling the collision timescale. Feature (ii) has potential importance for
the design of an efficient template for extracting the gravitational wave
signal from noise.

The close approximation \cite{pp}, based on the perturbation of a stationary
single black hole spacetime, has been extremely useful for understanding the
binary black hole problem by means of Cauchy evolution
\cite{price97,Campanelli98b,Campanelli98a,Campanelli98c,price99,price99l,price99l2,Campanelli99,Baker99a,Baker2000b}.
Our previous application of characteristic evolution to compute the retarded
solution for a white hole fission \cite{close1} is the starting point for the
present work. Under time reversal, the retarded solution for a white hole
fission is equivalent to an advanced solution for a black hole merger. In the
advanced solution, ingoing radiation is absorbed by the black holes but no
outgoing radiation is emitted. The advanced solution \cite{close1} provides
stage I of a two stage approach to the binary black hole problem
\cite{kyoto,compnull,fiss}. In stage II, treated here, we present the
physically appropriate retarded waveform for a black hole collision
corresponding to the retarded solution in which there is no ingoing
radiation from past null infinity ${\cal I}^-$ in the background Schwarzschild
spacetime.

These perturbative results furnish a reference point for the physical
understanding of a fully nonlinear version of the same approach
\cite{kyoto,compnull,fiss} which is being pursued using existing characteristic
evolution codes~\cite{high,reduced}. These results also complement the physical
picture obtained in this approximation by Cauchy evolution.

The advanced solution for a black hole merger is simpler to compute in a
characteristic approach because of the relative location of the two null
hypersurfaces on which boundary information is known. One of the null
hypersurfaces is the black hole event horizon ${\cal H}^+$, whose perturbation
corresponds to the close approximation of a binary black hole. In the advanced
problem, the other null hypersurface (in a conformally compactified
description) is future null infinity ${\cal I}^+$ where the outgoing radiation
vanishes. Since ${\cal H}^+$ and ${\cal I}^+$ formally intersect at future time
infinity $I^+$, they can be used to pose a double-null initial value
problem \cite{sachsdn,haywdn,fried81,rend} to evolve backward in time and
compute the exterior region of spacetime. (Potential difficulties in dealing
with $I^+$ are avoided by posing the problem on an ingoing null hypersurface
${\cal J}^+$ which approximates ${\cal I}^+$ and intersects ${\cal H}^+$ at a
late time when the perturbation has effectively died out.) On the other hand,
in the retarded problem, the other null hypersurface is past null infinity
${\cal I}^-$ where the ingoing radiation is required to vanish. Because ${\cal
H}^+$ and ${\cal I}^-$ are disjoint, there is no direct way to base a
characteristic initial value problem on these two hypersurfaces.

The advanced solution provides the ingoing radiation incident from
${\cal I}^-$. In the context of linear perturbation theory, the time
translation and reflection symmetries of the background provide a
mathematically well defined way to carry out stage II by using this
ingoing radiation to generate a ``source free'' {\em advanced minus
retarded} solution. A retarded solution is then produced by
superposition with the stage I solution. In this paper, we deal with
the subtleties produced in implementing this procedure
computationally.

Time reflection symmetry allows the merger of two black holes to be described
as the fission of a white hole.  It is convenient here for computational
purposes to formulate the binary black hole problem in this time reversed sense
as a white hole fission since the characteristic evolution then takes the
standard form of being carried out forward in retarded time. From a time
reversed point of view, the stage I solution is equivalent to the retarded
solution for a  white hole fission, with the physically appropriate
boundary conditions that radiation is emitted but that there is no ingoing
radiation from ${\cal I}^-$; and, under time reversal, the stage II solution is
equivalent to the advanced solution for a  white hole fission, with
the boundary condition that radiation from ${\cal I}^-$ is absorbed but no
outgoing radiation is emitted.

In Sec.~\ref{sec:background}, we review the background material upon which this
work is based \cite{close1}. We use a characteristic reformulation of the
Teukolsky equations \cite{Teukolsky73} governing the perturbative Weyl tensor.
Previous work dealt with a characteristic evolution algorithm for solving the
Teukolsky equation for the ingoing null component of the Weyl
tensor~\cite{close1}. In order to carry out Stage II, we extend this work to an
evolution algorithm for evolving the outgoing null component. Characteristic
data obtained for the nonlinear description of a binary black hole spacetime is
used to induce close approximation data for the perturbative solution
considered here.

In Sec.~\ref{sec:strategy}, we provide the technical details of our global
strategy for producing the retarded waveform from a black hole process. In
Sec.~\ref{sec:numerical}, we describe the numerical implementation and
testing of
the evolution algorithm. Because of difficult computational problems of a
global nature, we develop two independent evolution algorithms and carry out
extensive testing in order to demonstrate that the final waveform is free of
numerical artifacts.

In Sec.~\ref{sec:waveforms}, we present the properties of the retarded waveform
from a head-on black hole collision in the close approximation. It has been
suggested that the black hole merger waveform might have a distinct signature
that is roughly independent of the details of the initial conditions. Our
results do confirm the generally accepted view that the waveform has a
quasinormal ringdown stage characteristic of the final black hole mass (which is
here the background mass). However, we find that the features of the main
waveform are sensitive to the presence of ingoing radiation fields of the type
that would be present, for instance, in the evolution of time reflection
symmetric Cauchy data.

Notation and Conventions: We use a metric of signature $(-+++)$ and a
null tetrad with  normalization $l^a n_a=-m^a\bar{m}_a=-1$, so that
$g_{ab}=2(m_{(a}\bar{m}_{b)}-l_{(a} n_{b)})$. We introduce standard
angular coordinates $x^A=(\theta,\phi)$ to represent the unit
sphere metric $q_{AB}$ and set $q^{AB} = q^{(A}\bar q^{B)}$, where
$q^{AB}q_{BC}=\delta^A_C$, with $q^A=(1,i/\sin\theta)$. We use $q^A$
to define the $\eth$ operator with the convention $\eth f =q^A
\partial_A f$, for a spin-weight 0 function $f$.  Complex conjugation
is denoted with a ``bar'', e.g. $\Re(f) =(f+\bar f)/2$.  These
conventions lead to a different form of the perturbation equations from
that originally given by Teukolsky.

\section{Background formalism\label{sec:background}}

Various coordinates were used in Ref.~\cite{close1} to describe curvature
perturbations of a Schwarzschild background. The simplest global choice are
Israel coordinates~\cite{israel} in which the Schwarzschild metric takes the
form
\begin{equation}
   ds^2  = -W du^2-2dud\lambda  +r^2q_{AB}dx^Adx^B,
\label{eq:amet}
\end{equation}
where $u$ is an affine parameter along the ingoing null hypersurface
$r=2M$ that forms the white hole horizon, $\lambda$ is an affine
parameter along the outgoing null geodesics in the direction of
increasing $r=2M -\lambda u/(4M)$ and

\begin{equation}
      W=\frac{2\lambda^2}{\lambda u -8M^2}.
\end{equation}
(Affine freedom has been used to set $\lambda=0$ on the white hole
horizon, to set $u=0$ at the $r=2M$ bifurcation sphere between the
black hole and white hole horizons and to set $g^{ab}(\partial_a u)
\partial_b \lambda=-1$). Israel coordinates cover the entire Kruskal
manifold with nonsingular and simple analytic behavior. The exterior
quadrant is given by $u<0$, $\lambda>0$.

Other coordinates more suitable for certain numerical purposes are
outgoing Eddington-Finkelstein coordinates $(\tilde u,r)$, in which the
Schwarzschild metric is
\begin{equation}
   ds^2  = -(1-\frac{2 M}{r})d\tilde{u}^2-2d\tilde{u}dr+r^2d\Omega^2.
   \label{eq:outEF}
\end{equation}
Here $u=-Me^{-\frac{\tilde{u}}{4M}}$ and $\tilde{u}$ is a Bondi
retarded time coordinate related to the standard Schwarzschild time
coordinate $t$ by $\tilde{u}=t-r^*$,  with
$r^*=r+2M\log(\frac{r}{2M}-1)$. These outgoing Eddington-Finkelstein
coordinates patch the lower half of the Kruskal manifold composed of
the exterior quadrant and the quadrant above the initial singularity,
with $\tilde{u}\rightarrow\infty$ at the black hole horizon.

Israel coordinates induce the nonsingular complex null tetrad
\begin{eqnarray}
  l^a&=&-\nabla^a u = (\frac{\partial}{\partial\lambda})^a =
  \left[0,1,0,0\right],\nonumber\\
    n^a&=&(\frac{\partial}{\partial u})^a -
       \frac{W}{2} (\frac{\partial}{\partial\lambda})^a =
   \left[1,\frac{\lambda^2}{4Mr},0,0\right],\nonumber\\
   m^a&=&\frac{1}{{\sqrt 2} r} q^a,\nonumber\\
     \bar{m}^a&=&\frac{1}{{\sqrt 2} r} \bar q^a.
\label{NPtetrad}
\end{eqnarray}
where
\begin{equation}
         q^a = (\frac{\partial}{\partial\theta})^a +
        \, \frac{i}{\sin\theta}(\frac{\partial}{\partial\phi})^a=
          \left[0,0,q^A \right].
\end{equation}
The Teukolsky equations \cite{Teukolsky73} for
the vacuum perturbation of the outgoing null component of Weyl
curvature, $\psi_0=C_{abcd}l^a m^b l^c m^d$ (in Newman-Penrose notation
\cite{Newman62a}) and the ingoing null component
$\psi_4=C_{abcd}n^{a}\bar{m}^{b}n^{c}\bar{m}^{d}$ are
\begin{equation}
(L_0 + \frac {L^2}{2r^2}) \psi_0 = 0, ~~~{\rm and}~~~
(L_4 +\frac {L^2}{2r^2} ) \psi_4 = 0, \label{teukopsi}
\end{equation}
where
\begin{eqnarray}
  L_0&=&\frac{1}{4Mr}\lambda^2\partial^2_\lambda+
    \partial_u\partial_\lambda-\frac{5}{4Mr}u\partial_u
    \nonumber \\
   &-&\frac{1}{2Mr^2}\lambda(3M-4r)\partial_\lambda
  +\frac{5}{2 r M}, \nonumber \\
  L_4 &=&\frac{1}{4Mr}\lambda^2\partial^2_\lambda+
   \partial_u\partial_\lambda-\frac{1}{4Mr}u\partial_u
   \nonumber \\
   &-&\frac{7}{2r^2}\lambda\partial_\lambda
    -\frac{r^2 - 16 M^2 + 4 M r}{2 M r^3}
\label{teukops}
\end{eqnarray}
and $L^2=-\bar \eth \eth$ is the angular momentum squared operator.

The Weyl components $\psi_0$ and $\psi_4$ have spin-weight $s=2$ and
$s=-2$, respectively. It is useful to convert Eq's.~(\ref{teukopsi}) -
(\ref{teukops}) into spin-weight-zero equations by setting $\psi_0
=\eth^2 \Phi_0$ and $ \psi_4 =\bar \eth^2 \Phi_4$. Furthermore, the
computation of asymptotic behavior is enhanced by working with the
fields $F_0 = r^5\Phi_0$ and $F_4 = r\Phi_4$, which generically have
finite non-vanishing limits at ${\cal I}^+$ for asymptotically flat
perturbations.

Although the Weyl scalars formed from the $(l^a,n^a)$ null vectors are
non-singular in the full Kruskal manifold, they are not well adapted
for numerical accuracy. In order to compute radiation at late times
near ${\cal I}^+$ it is advantageous to consider a boosted tetrad
$({\tilde l}^a, {\tilde n}^a, m^a, \bar m^a )$, with ${\tilde
l}^a=-\nabla^a \tilde{u}$ and satisfying ${\tilde
l}^a {\tilde n}_a = -1$, which is adapted to Bondi time $\tilde u$ at
${\cal I}^+$ rather than the affine horizon time $u$.  We denote the
corresponding Weyl components by ${\tilde \psi}_0 = C_{abcd} {\tilde
l}^a m^b {\tilde l}^c m^d$ and ${\tilde \psi}_4 = C_{abcd} {\tilde n}^a
\bar m^b {\tilde n}^c \bar m^d$. In particular, the late time behavior
of the radiation waveform can be more accurately computed by an
evolution algorithm for the Weyl component $\tilde \psi_4$.

Setting $\tilde\psi_0 =\eth^2 \tilde\Phi_0$ and $\tilde \psi_4 =\bar
\eth^2 \tilde \Phi_4$, the Teukolsky equations for the fields $\tilde
F_0 =r^5\tilde \Phi_0$ and $\tilde F_4 =r\tilde \Phi_4$ reduce to the
spin-weight zero forms
\begin{eqnarray}
\label{eq:tildF_0}
           \left (D^2 + S_0 \right) \tilde F_0 &=& 0, \\
\label{eq:tildF_4}
               \left (D^2 + S_4 \right) \tilde F_4 &=& 0,
\end{eqnarray}
where
\begin{eqnarray}
    S_0 &=&  \frac{16M(r-3M)}{ur^2}\partial_\lambda - \frac{30 M}{r^3}
    + \frac{(6+\bar \eth \eth)} {\,r^2} \nonumber\\
    &=& -\frac{4(r-3M)}{ r^2} \partial_r - \frac{30 M}{r^3}
    + \frac{(6+\bar \eth \eth)} {\,r^2}
    \label{eq:S0}
\end{eqnarray}
and
\begin{eqnarray}
  S_4 &=&  -\frac{16M(r-3M)}{u r^2} \partial_\lambda - \frac {6 M}{r^3}
      + \frac{(2+\bar\eth \eth)}{\,r^2} \nonumber\\
   &=& \frac{4(r-3M)}{ r^2} \partial_r - \frac {6 M}{r^3}
      + \frac{(2+\bar\eth \eth)}{\,r^2} ,
      \label{eq:S4}
\end{eqnarray}
and
\begin{eqnarray}
       D^2  &=& -\frac{2 \lambda ( 16 M^2 - u \lambda)}{(8 M^2 - u \lambda)^2}
          \partial_\lambda -2 \partial_u \partial_\lambda
          \nonumber\\
         &-&\frac{2 \lambda^2}{8 M^2 - u \lambda} \partial_\lambda^2,
\label{D}
\end{eqnarray}
is the Laplacian defined by the metric $d\bar s^2 =
-(1-2M/r)d\tilde{u}^2-2d\tilde{u}dr$ induced by the
background on the 2-dimensional $(\tilde u,r)$ subspace.
Equations~(\ref{eq:S0}) and (\ref{eq:S4}) asymptote to one-dimensional
wave equations for solutions whose radial derivative falls off
uniformly as $O(1/r^2)$. The limit of $\tilde F_4$ determines the
outgoing gravitational radiation waveform while the limit of $\tilde F_0$
is related to the retarded quadrupole moment of the system. More
precisely, $\lim r\tilde \psi_4 =\partial_{\tilde u} \bar N$, where
$N$ is the standard definition of the Bondi news function.

In these variables, the deviation of Eq's.~(\ref{eq:tildF_0}) and
(\ref{eq:tildF_4}) from a 1-dimensional wave equation is independent of
time at a fixed $r$. Because $\tilde F_4 = (u/4M)^2 F_4$, where $F_4$
is regular throughout the Kruskal manifold (since it is
constructed with a regular basis), it follows that $\tilde F_4
\rightarrow 0$ as the black hole horizon is approached. This
facilitates an accurate long term evolution of the waveform using
Eq.~(\ref{eq:tildF_4}).

In contrast, $\tilde F_0 = (4M/u)^2 F_0$, so that $\tilde F_0$ is
singular on the black hole horizon, and thus a poor choice of variable
for long term evolution. The opposite signs of the coefficients of
$\partial_r$ in $S_0$ and $S_4$ are responsible for this behavior, as
can be seen by ignoring the remaining potential terms and freezing the
coefficient of $\partial_r$ at $r=2M$, so that Eqs.~(\ref{eq:tildF_0})
and (\ref{eq:tildF_4}) reduce to
\begin{eqnarray}
\label{eq:wave0}
  (2\partial_{\tilde u} -\partial_r -\frac{1}{M})\partial_r\tilde F_0 &=& 0, \\
\label{eq:wave4}
    (2\partial_{\tilde u} -\partial_r +\frac{1}{M})\partial_r\tilde F_4 &=& 0,
\end{eqnarray}
in terms of retarded Bondi coordinates. In this approximation, both of
these equations admit purely outgoing waves $\tilde F(\tilde u)$.
However, an ingoing $\tilde F_0$ wave has the exponentially singular
behavior
\begin{equation}
   \tilde F_0 = f(\tilde u +2r)e^{\frac {\tilde u - \tilde u_0}{2M}}
\label{eq:expgr}
\end{equation}
as an initial pulse $f(\tilde u_0 +2r)$ approaches the black hole horizon as
$\tilde u \rightarrow \infty$. An ingoing $\tilde F_4$ wave behaves in opposite
fashion and decays exponentially on approach to the black hole.

The different forms of Eqs.~(\ref{eq:tildF_0}) and (\ref{eq:tildF_4}) make it
clear that the Weyl component $\psi_0$ in the outgoing null direction $l^a$ and
the Weyl component $\psi_4$ in the ingoing null direction $n^a$ are not related
in a way which makes manifest the Schwarzschild time reflection symmetry ${\cal
T}$, defined by ${\cal T}(t,r,\theta,\phi)=(-t,r,\theta,\phi)$ in Schwarzschild
coordinates or by ${\cal T}(\tilde u,r,\theta,\phi)=(-\tilde v,r,\theta,\phi)$
in terms of Bondi retarded and advanced times $\tilde u =t-r^*$  and $\tilde v
=t+r^*$. The time reflection symmetry could be incorporated explicitly into the
pair of Teukolsky equations by introducing null tetrad vectors $L^a=\alpha l^a$
and $N^a=(1/\alpha) n^a$ satisfying ${\cal T}L^a=-N^a$. However, the explicit
form of the required boost,
\begin{equation}
      \alpha = -\frac {2M}{u} \sqrt{ \frac{2(r-2M)}{r}  }
             = -\frac {2M}{u} \sqrt{ \frac{-2\lambda u}{8M^2-\lambda u}  } ,
\end{equation}
makes it clear that such a time symmetric formulation would introduce
singular behavior at both the black and white hole horizons.

This time symmetric tetrad is useful for formulating the time reflection
properties of solutions of the Teukolsky equations using other tetrads. Let
$\Psi_4=C_{abcd}N^a \bar m^bN^c \bar m^d = \Psi(\tilde u,r,\theta,\phi)$  be a
perturbative solution for $\Psi_4$. Then the time reflection symmetry implies
that $\Psi_0=C_{abcd}L^a m^bL^cm^d = \bar \Psi(-\tilde v,r,\theta,\phi)$ is a
perturbative solution for $\Psi_0$. This correspondence maps a retarded
solution (no incoming radiation) for $\Psi_4$ into an advanced solution (no
outgoing radiation) for $\Psi_0$. In terms of the $\tilde l^a$ and $\tilde n^a$
Weyl components, the solution $\tilde \psi_4=\tilde \psi(\tilde
u,r,\theta,\phi)$ corresponds under time reflection to the solution
\begin{equation}
  \tilde \psi_0=\frac {4r^2}{(r-2M)^2}\bar \psi(-\tilde v,r,\theta,\phi).
\label{eq:trefl}
\end{equation}

\subsection{Close approximation data\label{sec:close}}

Data for the double-null formulation of the characteristic initial value
problem are given on a pair of intersecting null hypersurfaces. The data for a
fissioning white hole is posed on the ingoing null hypersurface ${\cal H}^-$
and an early outgoing null hypersurface ${\cal J}^-$, which extends from ${\cal
H}^-$ to ${\cal I}^+$. In Ref.~\cite{close1}, consistent double-null data for
$\psi_4$ for a fissioning white hole was obtained from the metric data for the
nonlinear version of the problem in terms of a spin-weight-2 field $J$
describing the inner conformal geometries of ${\cal H}^-$ and ${\cal
J}^-$ \cite{compnull}.

The underlying field $J$ for a white hole fission is provided by a conformal
model based upon an ingoing null hypersurface $\cal N$ emanating from a prolate
spheroid embedded in a flat space~\cite{ndata,asym}. Let $(\hat r, \theta,
\phi)$ be standard spherical coordinates for the inertial time slices $\hat t
=constant$ of Minkowski space. When the eccentricity of the spheroid vanishes,
the Minkowski null hypersurface reduces to the light cone from a sphere $\hat
t=0$, $\hat r= a$. The intrinsic geometry of ${\cal N}$ is described by a
degenerate rank-2 metric $\hat r^2 h_{AB}$ where $J=q^A q^B h_{AB}$ .  To
linear order in the eccentricity, the conformal model gives
\begin{equation}
     J(\hat t, \theta) = -\frac {a \sin^2\theta}{\hat t -a}.
\label{eq:closej}
\end{equation}

This conformal geometry of ${\cal N}$ models the conformal geometry of the
white hole horizon ${\cal H}^-$ in the close limit where the fission occurs in
the infinite future. The area coordinate $r$ of the geometry of ${\cal H}^-$
differs from the corresponding $\hat r$ of the geometry of ${\cal N}$ in order
to yield an asymptotically constant surface area for the initial white hole in
the infinite past. For vanishing eccentricity, this produces a Schwarzschild
horizon and a small eccentricity generates the close approximation.

The Raychaudhuri equation, which governs the expansion of a null hypersurface,
forces a difference in the affine parameters on ${\cal N}$ and ${\cal H}^-$.
With a suitable choice of affine scale, the relation between $\hat t$ and the
affine parameter $u$ on the white hole horizon is
\begin{eqnarray}
  \frac{d\hat t}{d u} &=& \Lambda (\hat \tau)
    =\frac {\hat \tau^2 (\hat \tau-1)^2}
      {(3 -5\hat \tau +\hat \tau^2)^2} \nonumber \\
      & \times& \bigg (\frac {(5-\sqrt{13}) -2\hat \tau}
                 {(5+\sqrt{13})  -2\hat \tau}\bigg )^{4/\sqrt{13}},
\label{eq:affine}
\end{eqnarray}
where $\hat \tau= \hat t -a$.  At early times Eq.~(\ref{eq:affine}) implies
$u\sim \hat t$ but as the Minkowski null cone pinches off at $\hat t=a$ the
corresponding affine time on the white hole horizon asymptotes to
$u\rightarrow \infty$. In terms of the inverted pair-of-pants picture for a
white hole fission, the pants legs are mapped to $u=\infty$, so that in the
close limit the individual white holes are mapped to future infinity along the
white hole horizon in the Kruskal manifold. The details are discussed
elsewhere in a treatment of fully nonlinear null data for the general two
black hole problem~\cite{compnull,asym}.

Close limit data for $J(u,\theta)$ on the white hole horizon is
determined by integrating Eq.~(\ref{eq:affine}) and substituting into
Eq.~(\ref{eq:closej}). In order to eliminate nonessential integration constants,
we set $u=0$ at the $r=2M$ bifurcation sphere. Then, up
to scale, the close data are determined by the single parameter
\begin{equation}
  \eta = -\frac { J|_{u=0} }{ u J|_{u=\infty} }= -\frac {1}{\hat \tau |_{u=0}},
   \label{eq:eta}
\end{equation}
where $\eta >0$ is a scale invariant parameter determining the yield of the
white hole fission. (In the time reversed scenario of a black hole collision,
$\eta$ is related to the inelasticity of the collision.) In the linear
approximation,
\begin{equation}
      \bar \psi_4 \approx {\frac{1}{2}} J_{,uu} .
      \label{eq:jtopsi}
\end{equation}

In this way, the conformal horizon model determines the close limit
data for $\psi_4$ on the horizon ${\cal H}^-$.
 By choosing ${\cal J}^-$ to be an early
outgoing null hypersurface approximating ${\cal I}^-$, we prescribe
data $\psi_4 =0$ on  ${\cal J}^-$, which is consistent with the
asymptotic falloff  $\psi_4 = O(u^{-3})$ along ${\cal H}^-$ implied by
Eq.~(\ref{eq:closej}). This approximates the condition that there be no
ingoing radiation from ${\cal I}^-$.

In terms of the spin-weight-zero potential $F_4$, the close data on
${\cal H}^-$ have the normalized form

\begin{equation}
  F_4 = -(\Lambda \partial_{\hat \tau})(\Lambda \partial_{\hat \tau})
            \frac{1}{\hat \tau},
\label{eq:closej2}
\end{equation}
after factoring out the $\ell=2$ angular dependence.

The time dependence of the close approximation data is quite mild when
expressed as a function of $\hat \tau$, as in Eq.~(\ref{eq:closej2}).
However, the relationship (\ref{eq:affine}) produces a sharp transition
region where the behavior of $\hat \tau (u)$ changes from the
asymptotic form  $d\hat \tau /du  \to 1$ as $\hat \tau \to - \infty$ to
$d\hat \tau /du  \to 0$ as $\hat \tau \to 0$~\cite{compnull}. For large
values of the parameter $\eta$, this produces sharply pulse shaped
data~\cite{close1}.

\section{Computing retarded waveforms from black holes\label{sec:strategy}}

\subsection{Removing the ingoing radiation\label{sec:global}}

In the time reversed scenario being pursued here, our strategy for computing
the retarded solution for a black hole collision is based upon the computation
of the advanced solution from a white hole fission. Stage I of the approach,
{\it i.e.} computation of the retarded solution for a white hole fission, has
already been carried out~\cite{close1}. Close approximation data for $\psi_4$
provided by the conformal horizon model was evolved to determine the stage I
solution $\psi_{4,I}$ throughout the exterior region of spacetime. In practice,
the evolution computes the corresponding spin-weight-0 potential $\tilde
F_{4,I}$, whose value on ${\cal I}^+$ determines the outgoing radiation field
$\tilde F_{4,I,OUT}$. In stage I of the white hole problem, there is no ingoing
radiation from ${\cal I}^-$. In stage II, the strategy for converting the
retarded solution to an advanced solution with no outgoing radiation is to
superpose a perturbative solution whose outgoing field equals $-\tilde
F_{4,I,OUT}$.

This procedure for converting from an advanced to a retarded field in a white
hole spacetime  differs from the standard procedure in a flat spacetime. The
retarded solution $\Phi_{RET}$ of the flat space wave equation $\Box \Phi = S$
with source $S$ may be converted into the advanced solution $\Phi_{ADV}$ by
superposing the source free (homogeneous) solution $H =\Phi_{ADV}-\Phi_{RET}$.
In the vacuum white hole (or black hole) problem, there is no source in the
exterior spacetime. The white hole horizon itself acts as the source of the
exterior solution. In this vein, specification of the ``source-free'' solution
which transforms a retarded white hole perturbation to an advanced white hole
perturbation requires specification of horizon data on ${\cal H}^+$ as well as
the outgoing data $-\tilde F_{4,I,OUT}$ at ${\cal I}^+$. This horizon data
must not introduce any features which change the physical interpretation of
the system as a white hole fission. For example, It is essential not to
superpose a black hole merger on the white hole fission, as would result in
the specification of a Cauchy problem with time symmetric data.

We carry out the conversion to an advanced solution by evolving backward in
time the double null problem with data $\tilde F_4 =0$ on the black hole
horizon ${\cal H}^+$ and data $\tilde F_4=-F_{4,I,OUT}$ at ${\cal I}^+$. This
data leaves the black hole horizon unchanged and its evolution determines the
analogue of the ``homogeneous solution'' $H$. The superposition $\tilde
F_{4,II} =\tilde F_{4,I}+H$ then produces the desired advanced solution for a
white hole fission with no outgoing radiation. The stage II solution leaves
the black hole horizon unchanged but perturbs the white hole, essentially
making it ring. In the time reversed black hole scenario, this is a physically
desirable effect because the binary black hole data provided by the conformal
horizon model has a power law decay in $u$, which corresponds to a purely
exponential decay in $\tilde u$ without quasinormal oscillations.  A priori
introduction of ringdown in the final black hole data would be artificial. In
stage II, the conversion from an advanced to retarded binary black hole
solution automatically introduces this ringdown in a natural way in addition to
removing the ingoing radiation.

\subsection{Producing the outgoing waveform\label{sec:asym}}

The Weyl components $\tilde \psi_0$ and $\tilde \psi_4$ form the major part of
the geometrical description of a Schwarzschild perturbation. The waveform at
${\cal I}^+$ is described directly by $\tilde \psi_4$. Stage II provides
indirect waveform information in terms of $\tilde \psi_0$, which can then be
used to obtain the waveform by means of differential relations between $\tilde
\psi_0$ and $\tilde \psi_4$. Asymptotically, at ${\cal I}^+$, these
differential relations can be obtained either from the vacuum Bianchi
identities \cite{nu} or from the asymptotic behavior of the metric
\cite{bondi,sachs,cbf}. Here we present the latter metric approach, since it
provides a complete asymptotic description of Schwarzschild perturbations.

In the $({\tilde u},r,x^A)$ Bondi coordinates with background Schwarzschild
metric (\ref{eq:outEF}), a vacuum perturbation near ${\cal I}^+$ perturbs the
metric in the asymptotically flat form~\cite{cbf}
\begin{eqnarray}
  \delta g_{a r}&=&\delta g_{AB}q^A \bar q^B =0 , \\
  \delta g_{\tilde u \tilde u} &=& \frac {2{\cal M}}
   {r}+O(\frac{1}{r^2}) ,\\
  \delta g_{\tilde u A}q^A &=&\frac {\bar \eth C}{4} +\frac{2 {\cal
      L}}{3r}
            + O(\frac{1}{r^2}) ,\\
  \delta g_{AB}q^A q^B &=& Cr+\frac {Q}{r}+O(\frac{1}{r^2}),
\end{eqnarray}
where $C$, $Q$, ${\cal M}$ and ${\cal L}$ are spin-weighted functions
of $(\tilde u,x^A)$. The spin-weight-2 functions $C$ and $Q$ comprise
the asymptotic part of the initial null data, in metric form, on a
constant $\tilde u$ hypersurface. In Bondi's terminology, the
spin-weight-0 function ${\cal M}$ is the (perturbed) mass aspect of the
system and the spin-weight 1 function ${\cal L}$ is the angular
momentum aspect. The time derivative of $\delta g_{AB}q^A q^B$ can be
determined from the evolution equations. The time derivative
$\partial_{\tilde u} C$ is the Bondi news function and requires a
radial integration to determine its value in terms of a boundary
condition on an inner worldtube. The time derivative of $Q$ can be
determined locally.  Similarly, the time derivatives of ${\cal M}$ and
${\cal L}$ can be determined locally from the components of Einstein's
equations representing conservation laws. In an asymptotic Bondi frame,
the relevant evolution-conservation equations are Eq's (7.19) - (7.22)
of Ref.~\cite{cbf}. They imply
\begin{eqnarray}
  \partial_{\tilde u} Q &=& \frac {\eth {\cal L}} {3}+\frac {{\cal M}
     C}{2} ,
\label {eq:qdot}\\
  \partial_{\tilde u} {\cal M} &=&\frac {\eth^2 \dot {\bar C}
        +\bar \eth^2 \dot C}{16}
\label {eq:mdot} ,\\
     \partial_{\tilde u}{\cal L} &=& \eth {\cal M}
         +\frac {\eth (\eth^2 \bar C-\bar \eth^2 C)}{16}.
\label {eq:ndot}
\end{eqnarray}

The asymptotic values of the Weyl components are obtained from the
metric according to
\begin{eqnarray}
  \tilde \psi^0_0 :&=&lim_{r\rightarrow \infty}r^5 \tilde \psi_0 =
     \frac{3}{2}Q \\
  \tilde \psi^0_4 :&=&lim_{r\rightarrow \infty} r\tilde \psi_4 =
     \frac{1}{4} \partial_{\tilde u}^2 \bar C.
\end{eqnarray}
This allows determination of  $\tilde \psi^0_4$ in terms of $\tilde
\psi^0_0$ by solving for $C$ in terms of $Q$. By taking 2 $\tilde
u$-derivatives of Eq.~(\ref{eq:qdot}) and using Eqs.~(\ref{eq:mdot})
and (\ref{eq:ndot}) to eliminate ${\cal M}$ and ${\cal L}$, we obtain
\begin{equation}
   \partial^3_{\tilde u} Q = \frac {1}{24}\eth^4 \partial_{\tilde u}
\bar C+
                 \frac{M}{2}\partial^2_{\tilde u} C.
\label{eq:Q3dot}
\end{equation}
Replacing $C$ and $Q$ by spin-weight 0 potentials defined by $C=\eth^2
c$ and $Q=\eth^2 q$, Eq.~(\ref{eq:Q3dot}) reduces to
\begin{equation}
   \partial^3_{\tilde u} q =
              \frac {1}{24}L^2 (L^2-2) \partial_{\tilde u} \bar c
                + \frac{M}{2} \partial^2_{\tilde u} c .
\label{eq:q3dot}
\end{equation}

When $c$ is real, the solutions of Eq.~(\ref{eq:q3dot}) for which $q=0$ are
solutions to the linearized Robinson-Trautman equation. Since the
Robinson-Trautman solutions are singular as $\tilde u \rightarrow -\infty$, the
requirement of nonsingular initial conditions leads to a one-to-one
correspondence between $c$ and $q$. The axisymmetric head-on collision of two
black holes corresponds to the case where $c$ is a real $(\ell=2, m=0)$
spherical harmonic. In that case, Eq.~(\ref{eq:q3dot}) gives
\begin{equation}
     \partial_{\tilde u} c= \frac{2}{M} e^{-2\tilde u/M}
         \int_{-\infty}^{\tilde u} d{\tilde t}
          e^{2\tilde t/M}\partial^3_{\tilde t} q(\tilde t).
\label{eq:cdot}
\end{equation}

When $c$ is complex, the solutions for $q=0$ correspond to a twisting
version of a Robinson-Trautman perturbation. In this case, the
imaginary part of $c$ goes exponentially to $\infty$ as $\tilde u
\rightarrow +\infty$ and the relation between $c$ and $q$ is more
complicated. The complex case will not be considered further here.

For the quadrupole case, Eq. (\ref{eq:q3dot}) determines the asymptotic relation at ${\cal I}^+$
\begin{equation}
  \frac{1}{6} \partial^4_{\tilde u} \tilde F_0 =
              \tilde F_4
                + \frac{M}{2} \partial_{\tilde u} \tilde F_4 ,
\label{eq:q4dot}
\end{equation}
which determines the waveform $\tilde \psi^0_4=\bar \eth^2 \tilde F^0_4$ at
${\cal I}^+$ in terms of the asymptotic field $\tilde \psi^0_0= \eth^2 \tilde
F^0_0$. This relation can also be expressed in the integral form
\begin{eqnarray}
      \tilde \psi^0_4 &=&\partial_{\tilde u} \bar N   \\
       N &=& \frac{1}{4}\partial_{\tilde u} C=\frac{1}{2M} e^{-2\tilde u/M}
            \int_{-\infty}^{\tilde u} d\tilde t e^{2\tilde t/M}
               \partial^3_{\tilde t} Q(\tilde t) \nonumber \\
              &=&\frac{1}{3M} e^{-2\tilde u/M}
            \int_{-\infty}^{\tilde u} d\tilde t e^{2\tilde t/M}
               \partial^3_{\tilde t} \tilde \psi^0_0 (\tilde t),
\end{eqnarray}
where $N$ is the Bondi news function,

\section{Numerical algorithms and accuracy\label{sec:numerical}}

The stability of the background Schwarzschild geometry and the insensitivity of
the Weyl tensor to gauge problems imply the absence of any theoretical
limitations to the accurate numerical evolution of the Teukolsky equation in
the exterior Kruskal quadrant, where the background spacetime is nonsingular.
However, there is a practical limit to how close a numerical evolution can
produce meaningful results near future time infinity $I^+$ or past time
infinity $I^-$. There is no nonsingular way to treat time infinity as a
boundary in a conformally compactified  manifold, as is possible for null
infinity. Thus time infinity can only be approached asymptotically and the
asymptotic values of fields can depend on the direction of approach. This
direction dependent limit at $I^+$ makes an evolution of $\psi_0$  difficult.

Numerical difficulties have led us to implement the Stage II evolution of
$\tilde F_0$ using two distinct approaches, one based upon a retarded time
foliation and the other on an advanced time foliation. Both approaches require
finding variables which allow for stable evolution of waves propagating inside
the peak of the Schwarzschild potential, which turns out to be difficult. The
retarded time algorithm uses an evolution variable which is adapted to the
falloff at $\scri^+$, and the advanced time algorithm uses an evolution
variable adapted to the falloff at $\scri^-$. Both approaches have similar
accuracy but the advanced time algorithm is more efficient at treating behavior
near ${\cal H}^+$ and the retarded time algorithm is more efficient near
$\scri^+$.

\subsection{Retarded time algorithm}

We base the retarded time algorithm  on the evolution variable
\begin{equation}
     \hat F_0 = (1-2M/r)^2\tilde F_0,
\end{equation}
which is finite on ${\cal H}^+$, as well as on $\scri^+$ where the waveform is
read off.  However, the factor $(1-2M/r)^2$, which makes $\hat F_0$ vanish
identically on ${\cal H}^-$, introduces a singularity at ${\cal H}^-$ in the
associated Teukolsky equation. In addition, $\hat F_0$ is infinite on
$\scri^-$.  We deal with these problems by initializing the double-null problem
on an ingoing null hypersurface ${\cal K}^-$ approximating ${\cal H}^-$ and on
an outgoing null hypersurface ${\cal J}^-$ approximating $\scri^-$.

In $(\tilde u, \tilde v)$ double null coordinates,
the Teukolsky equation for $\hat F_0$ ($\ell = 2$) is
\begin{eqnarray}
     \left (\partial_{\tilde u} \partial_{\tilde v} \right.&+&
     \frac {2 (1-2 M /r)}{r} \partial_{\tilde v} -
    \frac{2 M }{r^2} \partial_{\tilde u} \nonumber \\
    &+&
    \left. \frac{3 M (1- 2 M /r)}{2 r^3}
      \right ) \hat F_0 = 0.
\label{eq:teuk_hat_f0}
\end{eqnarray}
The code is implemented in coordinates $(\tilde u, \rho$), where
$\rho= \arctan(\tilde v /
\alpha)$ and $\alpha$ is an adjustable parameter (typically $\alpha = 40 M$).
These coordinates compactify ${\cal I}^+$ at $\rho=\pi /2$.
The Teukolsky equation takes the form
\begin{eqnarray}
   \partial_{\tilde u} \partial_{\rho} \hat F_0 &=&
    -{\frac {2 \left (r-2 M\right )}{r^2}}
    \partial_\rho \hat F_0
    \nonumber \\
     &+&{\frac {2\alpha\,\left(1+\tan^2 \rho \right)
     }{r^2}}\partial_{\tilde u}
      \hat F_0
                       \nonumber \\
   &-&\frac{3}{2}\,{\frac {\alpha\,\left (r-2 M \right )
    \left (1+ \tan^2 \rho \right )}{r^4}}\hat F_0
     ,
\label{eq:TeukEq_for_hatF0}
\end{eqnarray}
which is nonsingular in the region $ -\pi/2 < \rho < \pi/2$.   However,
compactification of $\scri^+$ using double null coordinates necessarily leads
to a loss of differentiability at ${\cal I}^+$ \cite{berndtub}. The
coefficients in Eq. (\ref{eq:TeukEq_for_hatF0}) have finite limits at
$\scri^+$, but the  $\partial_{\tilde u} \hat F_0$ coefficient is not
differentiable at $\scri^+$. In $(\tilde u,l=1/r)$ coordinates $\hat F_0$ can
have analytic dependence on $l$ at ${\cal I}^+$ but the transformation to
$(\tilde u, \rho)$ coordinates introduces logarithmic behavior so that $\hat
F_0$ is not twice differentiable with respect to $\rho$ at $\scri^+$. Our tests
show  this still allows the computation of $\hat F_0$ on $\scri^+$ to 2nd order
accuracy in grid size. Similar logarithmic behavior exists at $\scri^+$ in our
original Stage I algorithm \cite{close1}.

We implement Eq. (\ref{eq:TeukEq_for_hatF0}) as the first differential order system
\begin{eqnarray}
   \partial_{\tilde u} \hat F_{0}
   &=& g(\tilde u, \rho)
              - \frac{2(r-2M)}{r^2}\hat F_{0}
               \label{eq:hat_f0_time} \\
       \partial_{\rho} g
       &=&
    {\frac {2\alpha\,\left(1+\tan^2 \rho \right )}{r^2}} g
    \nonumber \\
   &-&\frac {\alpha\,\left (1+\tan^2 \rho \right )}{2\,r^4} \nonumber \\
   &\times& \left (r-2M \right )\left (3M+2\,r\right ) \hat F_0 
   , \label{eq:hat_f0_hyper}
\end{eqnarray}
where $g$ is introduced as an auxiliary variable.

The numerical $(\tilde u,\rho)$ grid consists of points on the $\tilde
u = const$ foliation. For reasons of economy and accuracy, the grid points
are uniformly spaced in $\tilde v$ in the region of compact support
where the initial data is non-zero and are uniformly spaced in $\rho$
in the outer region where the initial data vanishes.
We solve Eqs. (\ref{eq:hat_f0_time}) and (\ref{eq:hat_f0_hyper}) using a 2nd
order accurate Runge-Kutta scheme.

\subsection{Advanced time algorithm}

We base the advanced time algorithm on the evolution variable
\begin{equation}
      \breve{F}_0= (1 - 2 M/r)^2 r \tilde \Phi_0 =
       \frac{1}{r^4} \hat F_0,
\end{equation}
which is finite on ${\cal I}^-$ and on ${\cal H}^+$.  Since $\breve{F}_0$
automatically vanishes on ${\cal H}^-$, we must again initialize on an ingoing
null hypersurface ${\cal K}^-$ approximating ${\cal H}^-$. In addition,
$\breve{F}_0$ vanishes on ${\cal I}^+$. Consequently, the signal on ${\cal
I}^+$ must be obtained by transforming $\breve{F}_0$ to $\hat{F}_0$ on
successively farther out  $\tilde v=const$ slices  and extrapolating to $\tilde
v = \infty$. Although the advanced time algorithm can be initialized on
$\scri^-$ this gives no distinct advantage in computing the waveform at
$\scri^+$ and, in order to facilitate comparison with the retarded time
algorithm, we initialize on an outgoing null hypersurface ${\cal J}^-$ which
approximates $\scri^-$.

The advanced time evolution variable $\breve{F}_0$ satisfies
\begin{eqnarray}
  \bigg( \partial_{\tilde u} \partial_{\tilde v} &+&
  \frac{2 (r - 3 M)}{r^2} \partial_{\tilde u}  \nonumber \\ &+&
   \frac{(1- 2M/r)(2 r+ 3M)}{2 r^3} \bigg) \breve{F}_0 = 0
\label{eq:teuk_for_brev_f0}.
\end{eqnarray}
This is precisely the equation for $\tilde F_4$ under the transformation
$\tilde v \leftrightarrow - \tilde u$. (Under time reflection, $\breve F_0 = 4
{\cal T} \tilde F_4$). Thus the advanced time algorithm is equivalent to an
algorithm for evolving $\tilde F_4$ backward in time.

The evolution proceeds along $\tilde v = const$ null hypersurfaces, which are
compactified by the coordinate transformation $\tilde u = \alpha  \tan \mu$,
where  $-\pi/2 \le \mu \le \pi/2$ in the range from  ${\cal I}^-$ to $\cal
H^+$, and $\alpha$ is again an adjustable parameter (typically $\alpha = 200 M$
in the advanced time case). In these coordinates,
Eq. (\ref{eq:teuk_for_brev_f0}) becomes
\begin{eqnarray}
  \bigg (\partial_{\mu} \partial_{\tilde v} &+&
  \frac{2 (r - 3 M)}{r^2} \partial_{\mu}  \nonumber \\ &+&
   \frac{\alpha(1-2 M/r)(2 r + 3 M)}{2 r^3 \cos^2 \mu} \bigg ) \breve{F}_0 = 0.
\label{eq:com_TeukEq_for_breveF0}
\end{eqnarray}
This equation is non-singular in the region $-\pi/2 < \mu \le \pi/2$, but the
coefficient of $\breve F_0$  is not differentiable on $\scri^-$ ($\mu =
-\pi/2$).
On an ingoing null hypersurface, $\breve F_0$ can be expressed as a power
series in $l=1/r$.  This implies that the associated series in $\mu$ contains
logarithmic terms of the form $\mu (1 + \mu \log \mu)$, and is not twice
differentiable (with respect  to $\mu$) on $\scri^-$. Our tests indicate that
the code is 2nd order  convergent despite this mild lack of differentiability.

We implement Eq. (\ref{eq:com_TeukEq_for_breveF0}) in the first differential
order form
\begin{eqnarray}
  \partial_\mu \breve{F}_0 &=& \breve g \label{eq:brev_f0_hyper}\\
  \partial_{\tilde v} \breve g &=& -\frac {2 (r - 3 M)\breve g}{r^2}
    \nonumber\\ &-& \frac {\alpha (r-2 M)(2 r + 3 M)\breve{F}_0}
    {2 r^4 \cos^2 \mu}   , \label{eq:brev_f0_time}
\end{eqnarray}
where we have introduced the auxiliary variable $\breve g$.
The numerical $(\mu,\tilde v)$ grid consists of points on the $\tilde
v = const$ foliation equidistant in $\mu$.
The hypersurface equation (\ref{eq:brev_f0_hyper}) is solved by 2nd order
accurate integration and Eq. (\ref{eq:brev_f0_time}) is solved by a 2nd order
accurate Runge-Kutta scheme.

\subsection{Extracting the Retarded Waveform\label{sec:exret}}

The retarded waveform $\tilde F_4$ on $\scri^+$ is obtained from $\tilde F_0$
by means of Eq. (\ref{eq:q4dot}).  Given the values of $\tilde F_0$, we
integrate Eq. (\ref{eq:q4dot}) by the midpoint rule to obtain $\tilde F_4$,
using the initial condition that $\tilde F_4 = 0$.

The retarded algorithm provides $\tilde F_0$ on $\scri^+$ directly. In the
advanced algorithm, where $\scri^+$ is not in the evolution domain, we
extrapolate the waveform at $\scri^+$ from a quadratic polynomial based upon
approximate values of $\tilde F_4$ obtained by applying Eq. (\ref{eq:q4dot}) on
three ingoing null hypersurfaces approximating $\scri^+$ (where we substitute
$r^4 \breve F_0$ for $\tilde F_0$).  The three null hypersurfaces consist of
the advanced time slices $\tilde v_f$, $\tilde v_f/2$, and $\tilde v_f/3$,
where (typically) $\tilde v_f = 4000 M$.

\subsection{Initializing the Stage II evolution\label{sec:init_data}}

Initial data for the stage II evolution, i.e. the value of $\tilde \psi_0$ on
${\cal J}^-$, is obtained by a time reflection of the retarded Stage I solution
for $\tilde \psi_4$ on ${\cal J}^+$.  Technical problems arise here because the
stage I evolution does not provide accurate data at late times on ${\cal J}^+$
\cite{close1}. We
deal with this by smoothing the Stage II initial data to zero outside a region
of compact support where the data would otherwise be small. This modification
of the initial data  also guarantees consistency with the boundary data $\psi_0
=0$ on ${\cal K}^-$, which approximates the boundary condition that the
perturbation vanishes on ${\cal H}^-$.

The relation $\tilde \psi^0_4 =\partial_{\tilde u} \bar N$ on ${\cal I}^+$
requires that the integral
\begin{equation}
    \int_{-\infty}^{\infty}d\tilde u \tilde \psi^0_4=\bar N|_{-\infty}^{\infty}
\end{equation}
vanish since the news function $N$ must vanish at either limit in order that
the radiated mass loss be finite. By time reflection, it follows that the
integral
\begin{equation}
    \int_{-\infty}^{\infty}d\tilde v \breve F_0 =0
\label{eq:zeroint}
\end{equation}
on ${\cal I}^-$. However, artificially  truncating the initial data
to compact support causes this integral to be non-zero, although small. This
leads to the following artificial behavior of $\tilde F_0$ at early
retarded times.

This early time behavior of $\breve F_0$ is  conveniently described in
coordinates $(\tilde v, l=1/r)$ which compactify ${\cal I}^-$ at $l=0$. In
these coordinates, the evolution Eq. (\ref{eq:com_TeukEq_for_breveF0}) becomes
\begin{eqnarray}
        \bigg[ \partial_{l} \partial_{\tilde v} &+& (l^3 M - \frac{1}{2} l^2)
         \partial_l^2 +
          (l - 3 l^2 M) \partial_l \nonumber \\ &+& (2 + 3 M l) \bigg] \breve F_0 =0.
\label{eq:breve_f0_l}
\end{eqnarray}
Consider initial data on $\scri^-$ with compact support in the region
$\tilde v_a < \tilde v < \tilde v_b$. Integration of Eq. (\ref{eq:breve_f0_l})
along $\scri^-$ from $\tilde v = \tilde v_a$ to $\tilde v$ yields
$$
   \partial_{l} \breve F_0  = -2 \int_{\tilde v_a}^{\tilde v} d\tilde v
                  \breve F_0 d\tilde v = a(\tilde v),
$$
where according to the above argument $a(\tilde v)$ should vanish for $\tilde
v\ge \tilde v_b$. However, our numerical procedure for truncating the data
leads to a small value of  $a(\tilde v_b)$. Hence, for $\tilde v \ge \tilde
v_b$, on any ingoing null hypersurface, $\breve F_0$ has the asymptotic
behavior $\breve F_0(\tilde v, l) = a(\tilde v_b) l + O(l^2) $. When
re-expressed in $(\tilde u, \tilde v)$ double null coordinates, this implies
$\tilde F_0$ has the asymptotic behavior $\tilde F_0(\tilde u, \tilde v) =
a(\tilde v_b) \tilde u^3 + O(\tilde u^2)$. In numerical simulations, this
produces artificial cubic polynomial dependence of $\tilde F_0$ at early times
on ${\cal I}^+$.  Since this cubic behavior is removed by taking four
$\tilde u$-derivatives  it does not affect the waveform $\tilde F_4$
 calculated from $\tilde F_0$ using Eq. (\ref{eq:q4dot}).

\begin{figure}[tb]
 \centerline{\epsfxsize=2.75in\epsfbox{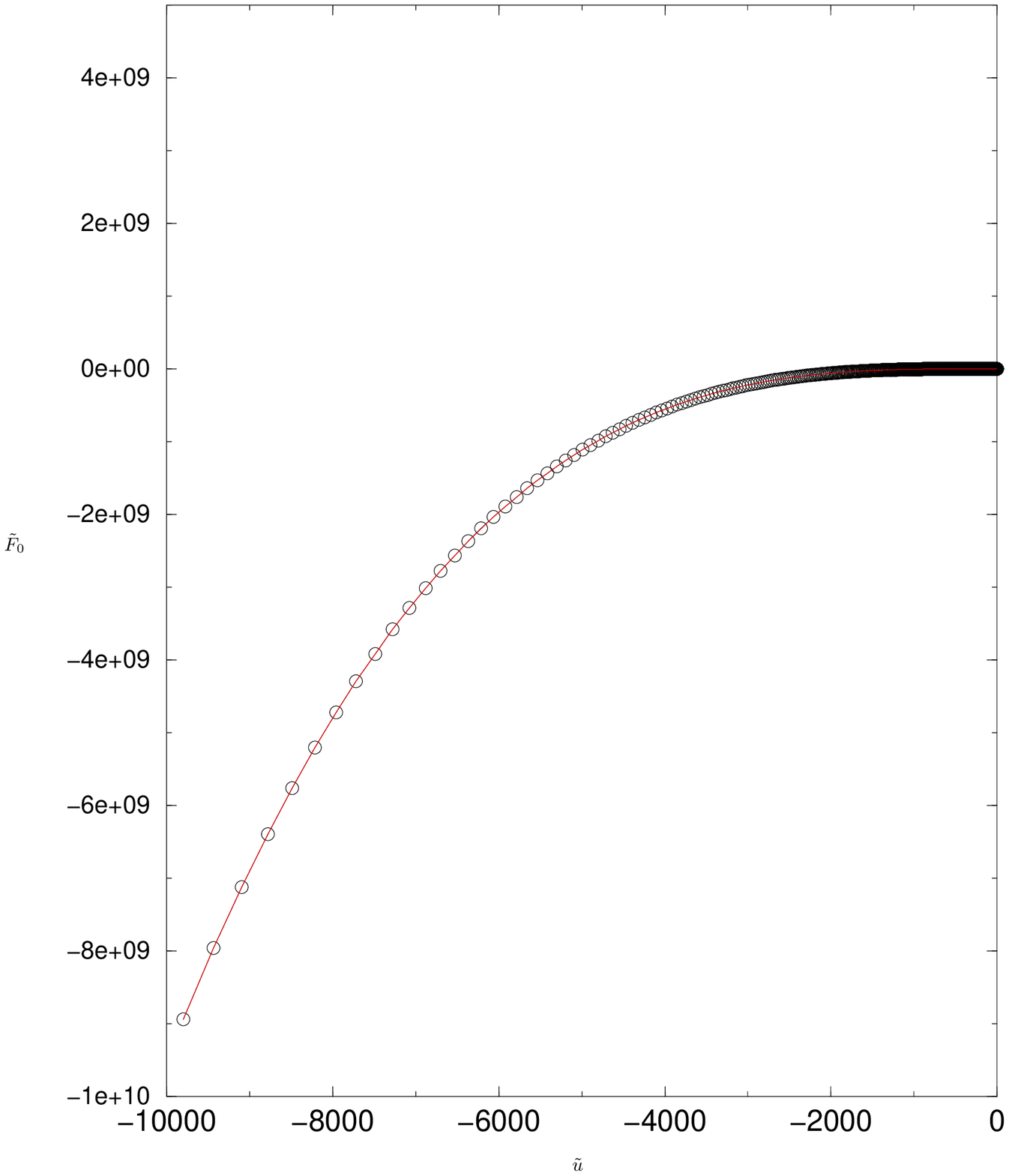}}
 \caption{Signal $\tilde F_0(\tilde u)$ produced on $\tilde v=1000M$ by an
 initial compact pulse. The solid curve indicates a fit to a cubic polynomial. }
\label{fig:early}
\end{figure}

\begin{figure}[tb]
 \centerline{\epsfxsize=2.75in\epsfbox{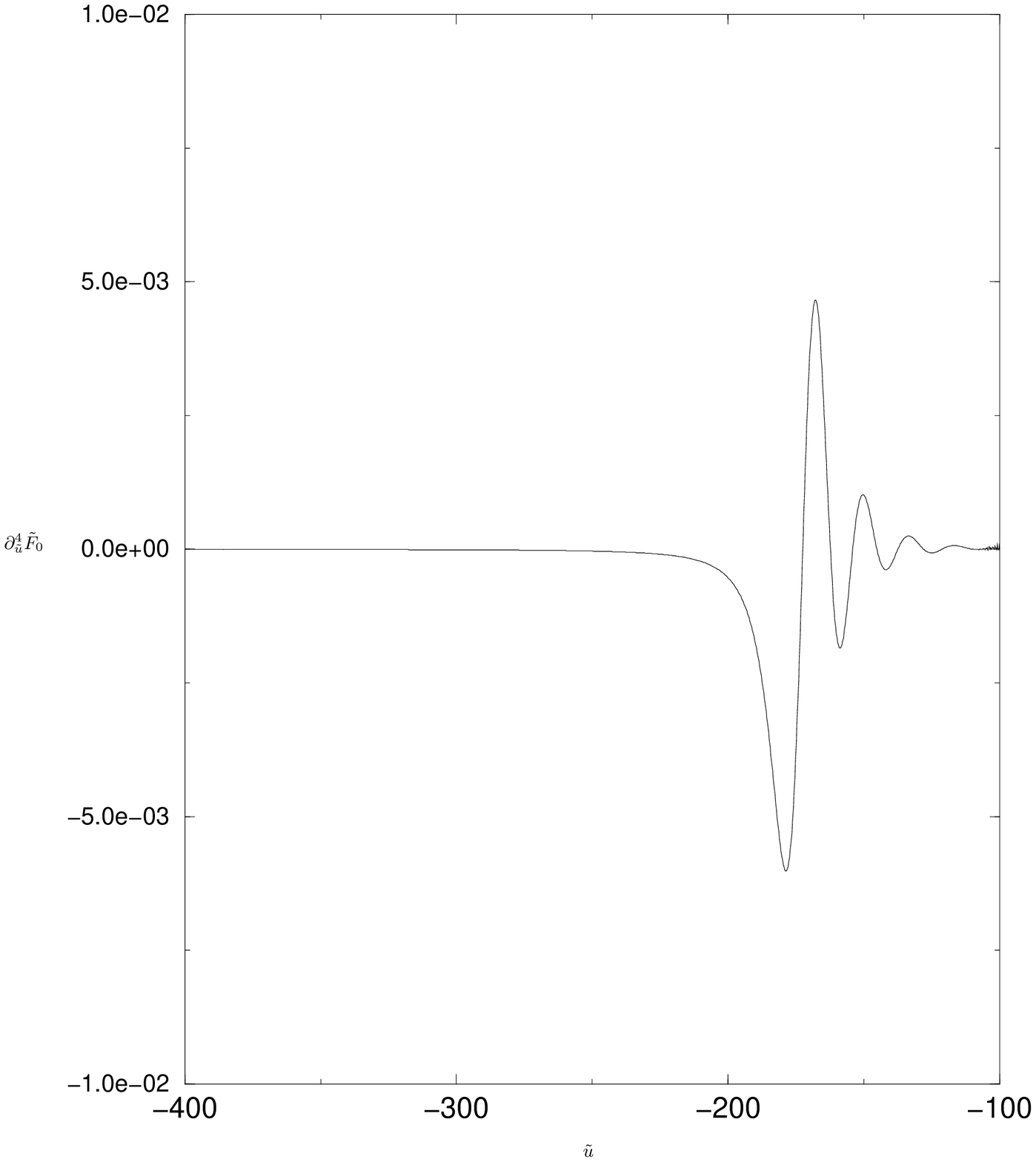}}
 \caption{The 4th $\tilde u$-derivative of Fig.~\ref{fig:early} .}
\label{fig:earlyderiv}
\end{figure}

We simulated this artificial early time behavior by using the
advanced time algorithm to propagate an extremely narrow initial
compact support pulse on $\cal I^-$ of the form  $\breve F_0 =
\left[(\tilde v - \tilde v_1)(\tilde v_2 - \tilde v)\right ]^4$
for $\tilde v_1 < \tilde v < \tilde v_2$, and $\breve F_0 = 0$
otherwise, where $\tilde v_2 - \tilde v_1  = .1M$.
Figure~\ref{fig:early}
plots the resulting signal $\tilde F_0$ on the  ingoing null
hypersurface $\tilde v = 1000 M$. (By looking at the signal on
finite $\tilde v$ we avoid problems associated with spatial
infinity $i^0$.) The figure shows that the early time behavior
fits a cubic polynomial in $\tilde u$. This cubic polynomial
dominates at later times and masks physical effects such as
quasinormal ringdown. However, Fig.~\ref{fig:earlyderiv}
shows that after taking four $\tilde u$-derivatives the ringdown
is clearly observed and there is little trace of the secular
behavior. Note the drastic difference between the scales of the
Figs.~\ref{fig:early} and \ref{fig:earlyderiv}.

In computing waveforms from close approximation data, because the initial
surface ${\cal J}^-$ is at finite $\tilde u$ the artificial early time behavior
in $\tilde F_0$ is described by a quartic rather than cubic polynomial in
$\tilde u$ but the 4th order coefficient is small for reasonably early start
times.  We performed the following experiments to measure the sensitivity of
the Stage II signal to the location of ${\cal K}^-$ and ${\cal J}^-$. In these
experiments we used a Stage I run to obtain close approximation initial data
with $\eta=1410$ for $\tilde F_0$  on $\cal J^-$ and measured the signal
propagated to $\scri^+$. Except when computational expense would have been
prohibitive, the experiments were carried out with both the retarded time and
advanced time evolution algorithms with virtually equivalent results.

\begin{figure}[tb]
 \centerline{\epsfxsize=2.75in\epsfbox{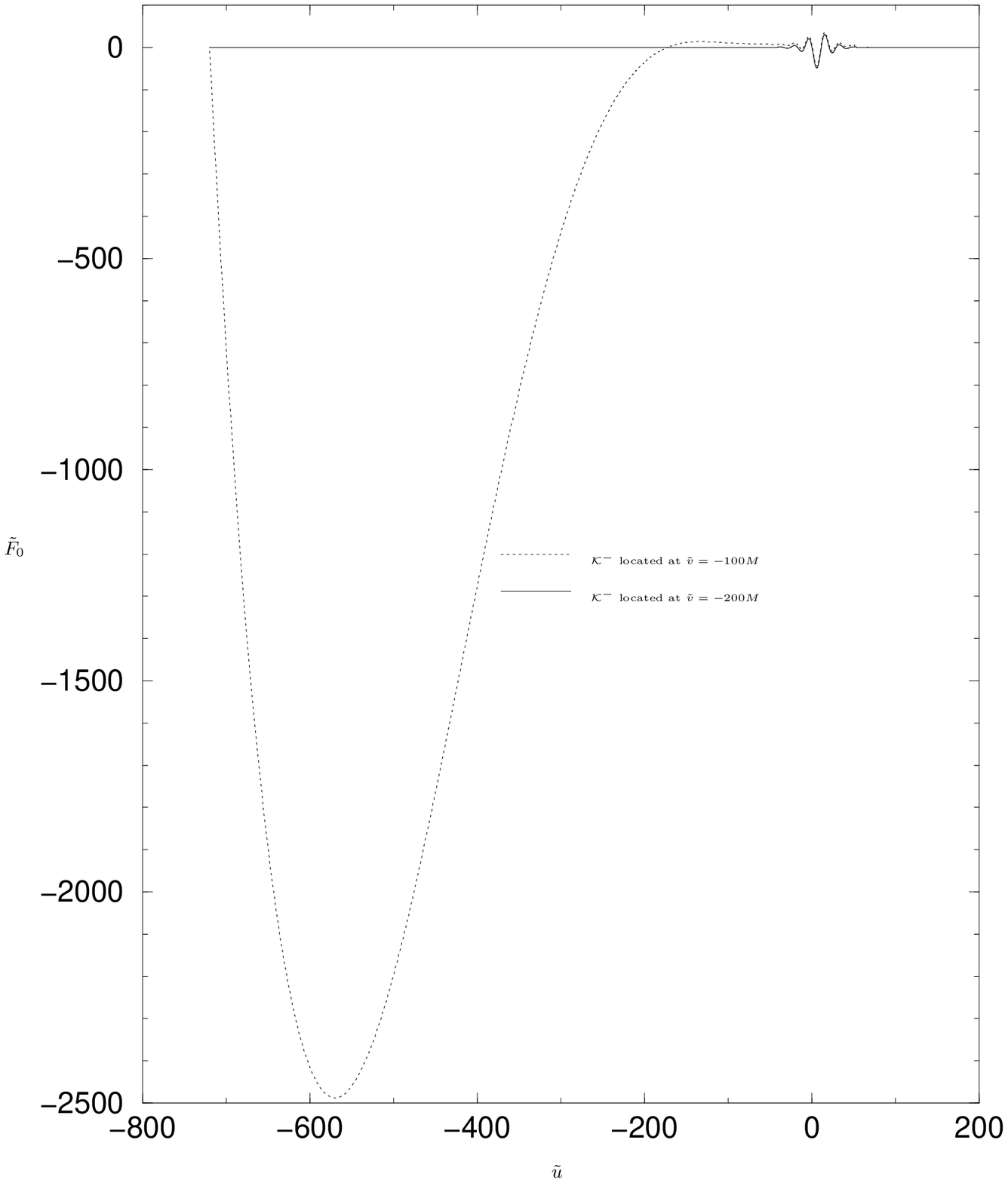}}
 \caption{Comparison of the behavior of $\tilde F_0$ for two locations
of ${\cal K}^-$. }
\label{fig:trunc}
\end{figure}

In the first set of experiments, ${\cal K}^-$ was placed at
$\tilde v=-200M$ and $\tilde v = -100M$, with the initial
hypersurface ${\cal J}^-$  located at $\tilde u = -720 M$ in both
cases. For $\tilde v=-100 M$ the initial data on ${\cal J}^-$
implied by the time reflection of the Stage I evolution was
approximately 5 orders of magnitude  smaller at ${\cal K}^-$ than
at its peak (at $\tilde v \approx 0$); and for $\tilde v=-200M$,
the size at ${\cal K}^-$ was 9  orders of magnitude smaller than
at the peak. The Stage II approximation of smoothing this data to
zero on ${\cal K}^-$ violates the integral condition
(\ref{eq:zeroint}) and produces the early time polynomial behavior
shown in the plots of $\tilde F_0(\tilde u)$ at ${\cal I}^+$ in
Fig.~\ref{fig:trunc}.
For both locations of ${\cal K}^-$ the early time behavior is well
fit by a 4th order polynomial but the size of the polynomial is
about $10^7$ times  larger for the farther out location of ${\cal
K}^-$.

\begin{figure}[tb]
 \centerline{\epsfxsize=2.75in\epsfbox{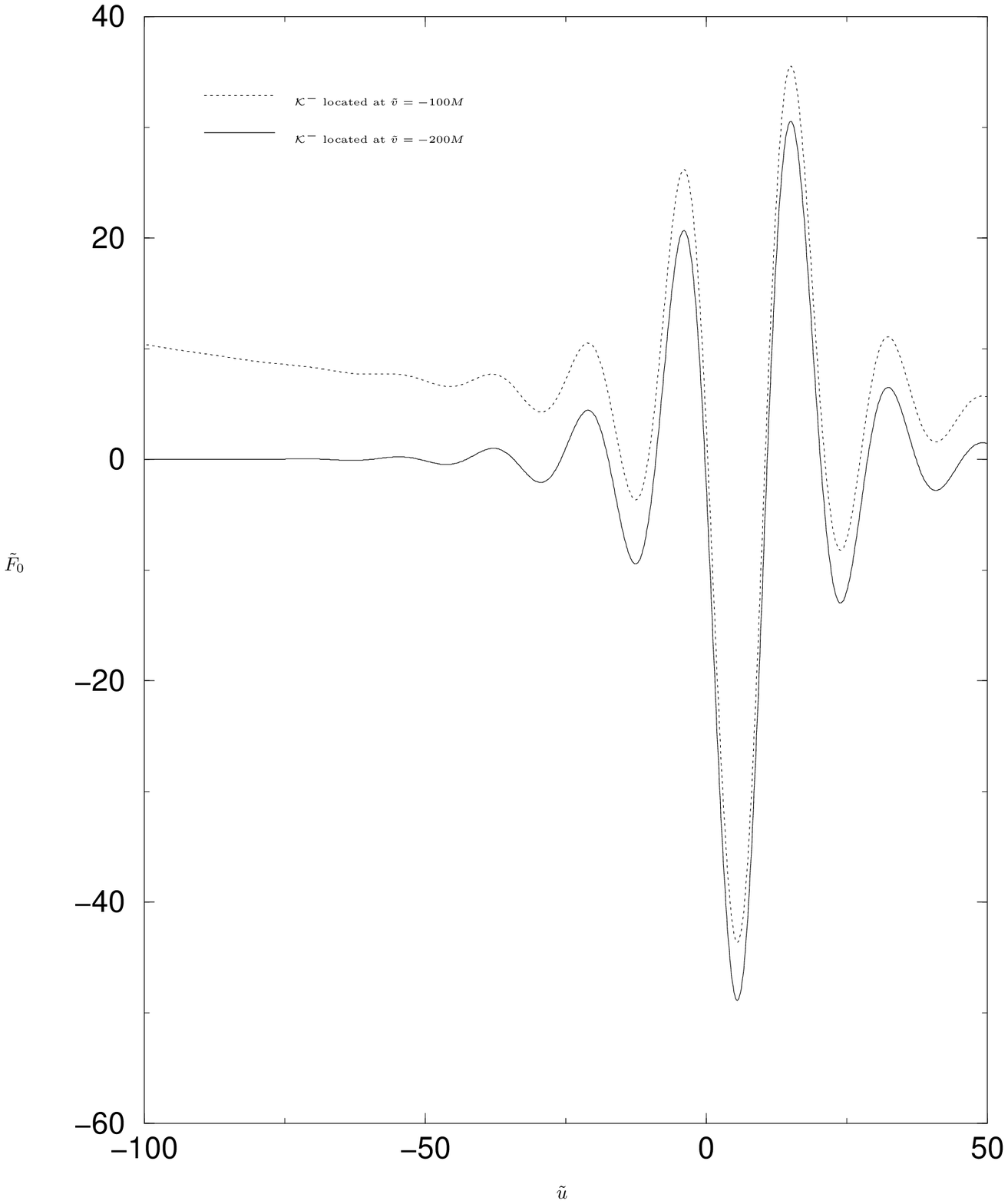}}
 \caption{Blowup of Fig.~\ref{fig:trunc} in the region of the genuine signal.}
 \label{fig:trunc_blowup}
\end{figure}

Figure~\ref{fig:trunc_blowup}
is a magnified view of Fig.~\ref{fig:trunc}
showing how the early time quartic polynomial behavior contaminates the genuine
signal at later times.

\begin{figure}[tb]
 \centerline{\epsfxsize=2.75in\epsfbox{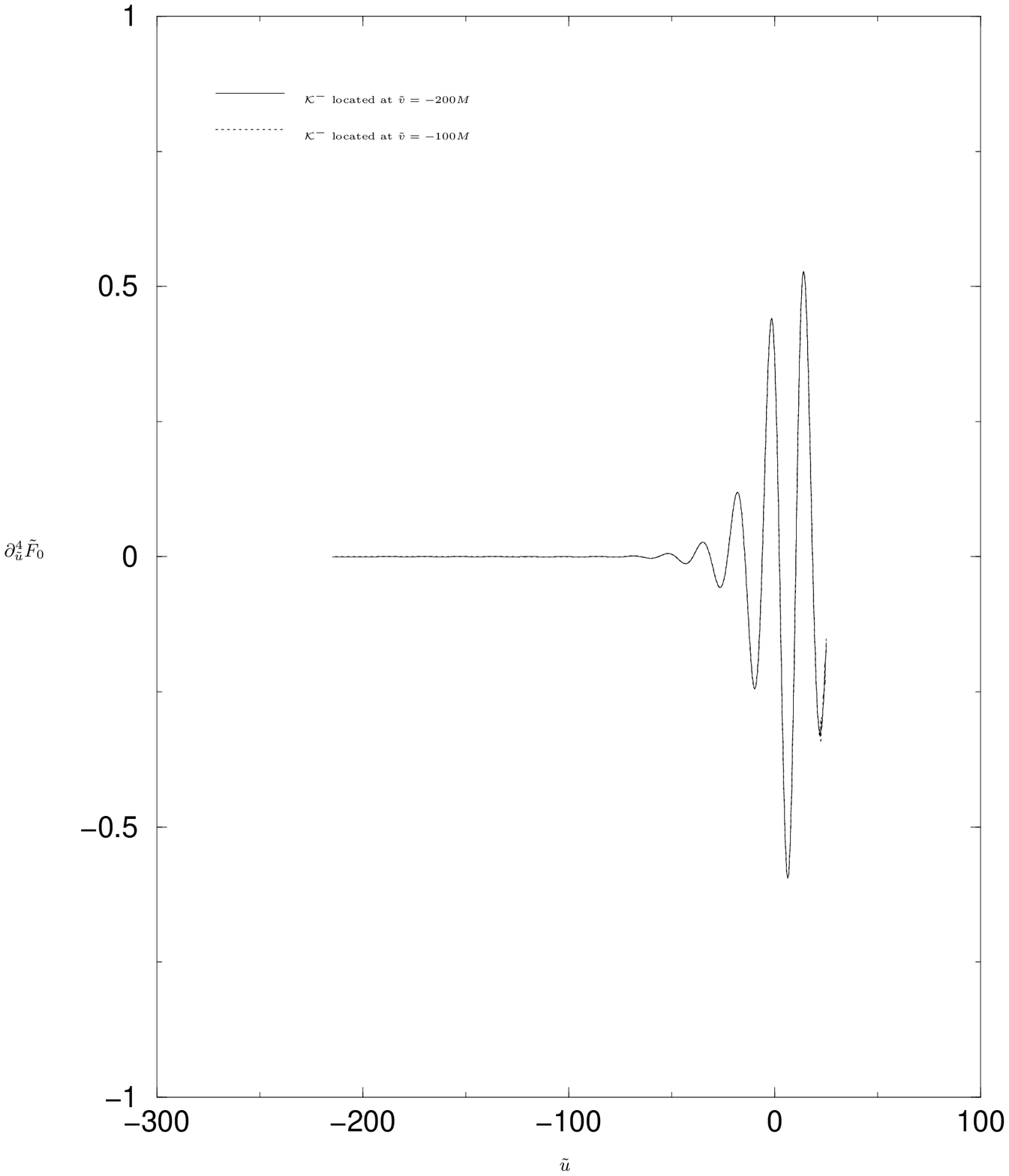}}
 \caption{The 4th $\tilde u$-derivative of Fig.~\ref{fig:trunc} shows
          independence of the location of ${\cal K}^-$. The evolution is terminated when
          late time noise begins to contaminate the results. (See Sec.~\ref{sec:exp}).}
 \label{fig:trunc_deriv}
\end{figure}

Figure~\ref{fig:trunc_deriv}
shows that the two signals agree after taking four $\tilde
u$-derivatives.

\begin{figure}[tb]
 \centerline{\epsfxsize=2.75in\epsfbox{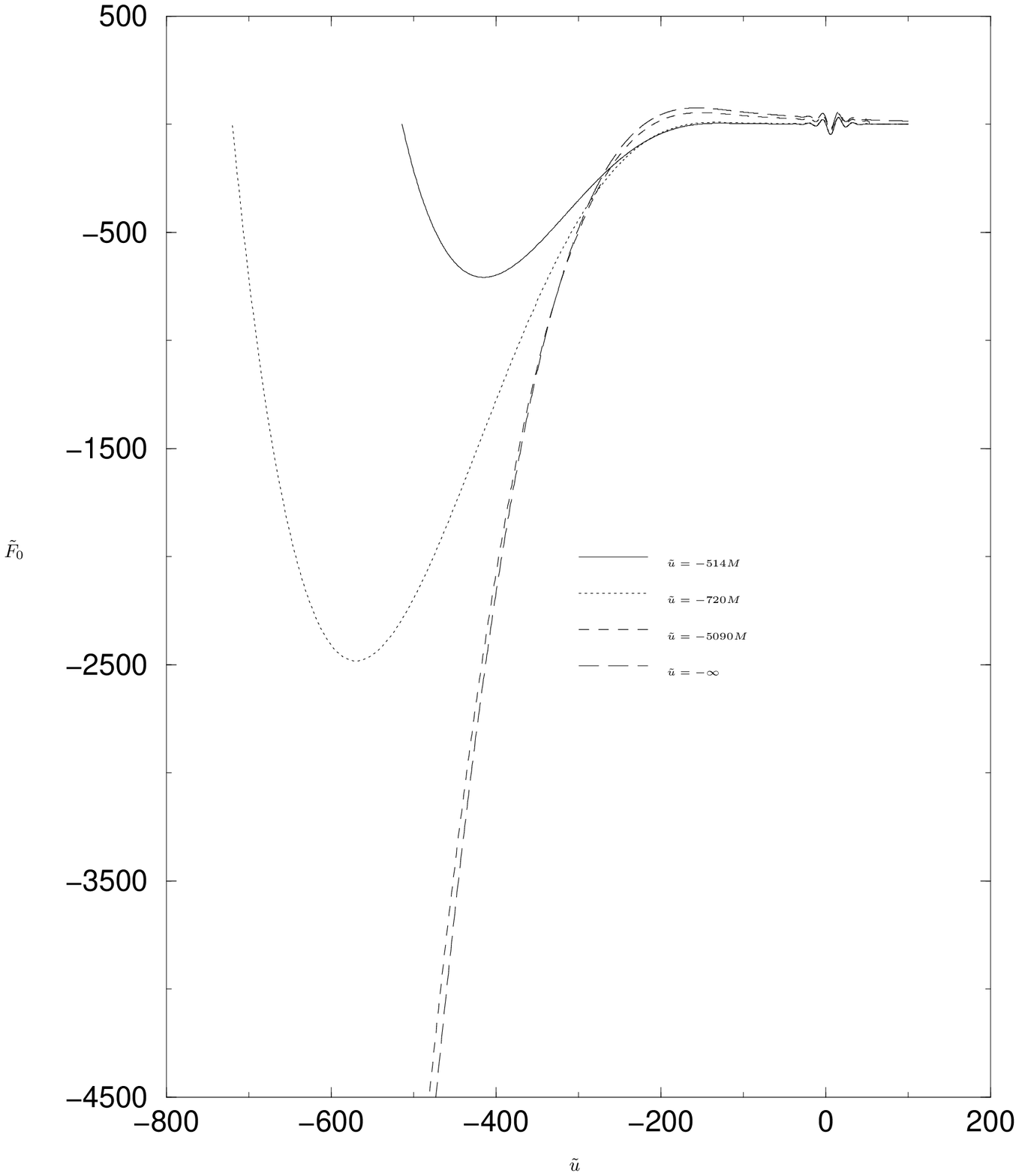}}
 \caption{Comparison of early time behavior for ${\cal J}^-$
          located at four different retarded times.}
 \label{fig:comp_start}
\end{figure}

In the second set of experiments, ${\cal J}^-$ was located at
$\tilde u = -540M$, $\tilde u = -720M $, $\tilde u = -5090 M$, and
$\tilde u = -\infty$, with ${\cal K}^-$ located at $\tilde v =
-100 M$ (to accentuate the early time polynomial behavior).
Figure~\ref{fig:comp_start},
which plots the resulting signal $\tilde F_0(\tilde u)$ on $\tilde
v = 1000M$, shows that the polynomial behavior becomes more
pronounced as ${\cal J}^-$ approaches ${\cal I}^-$.

\begin{figure}[tb]
  \centerline{\epsfxsize=2.75in\epsfbox{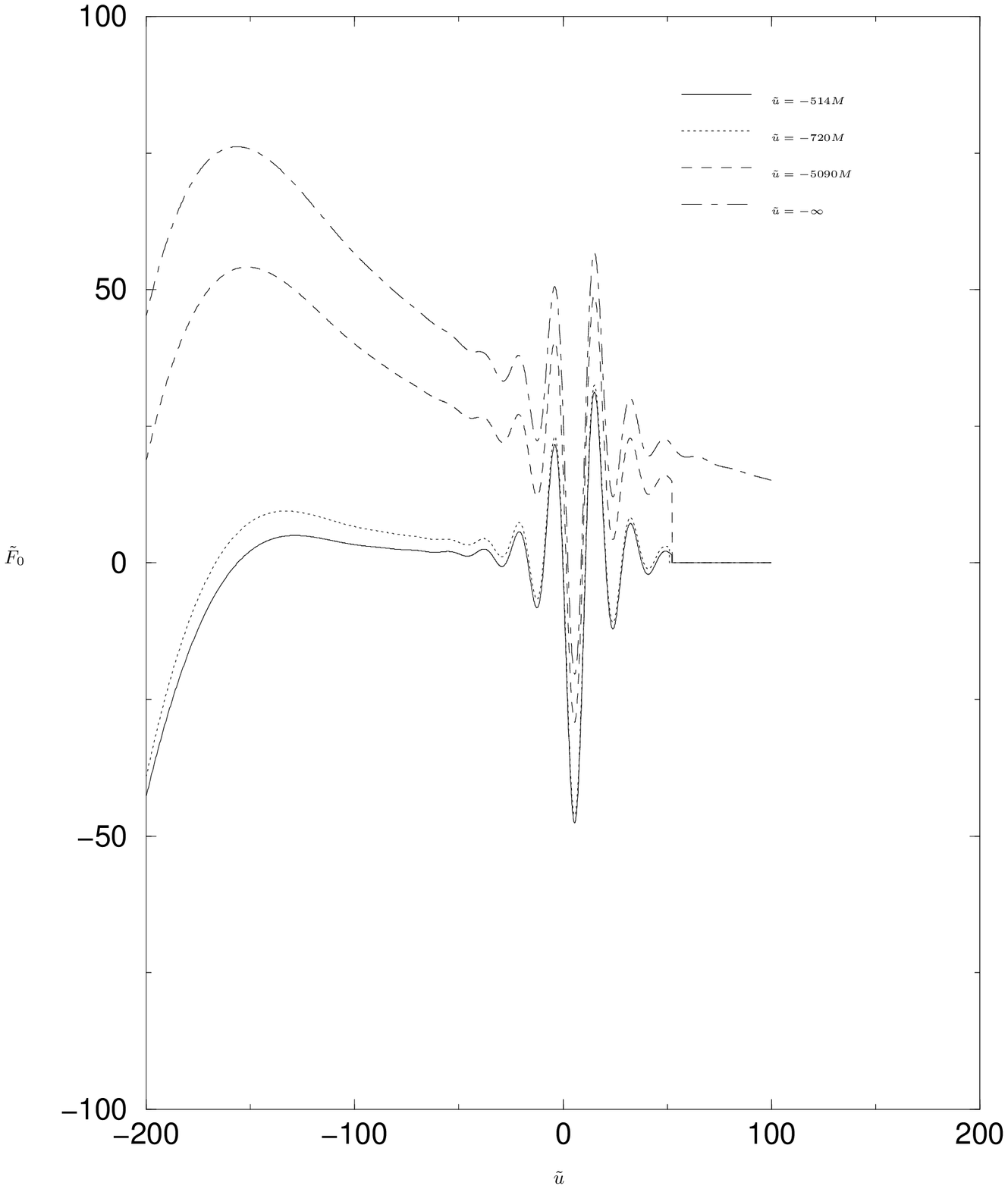}}
  \caption{Blowup of Fig.~\ref{fig:comp_start}. }
  \label{fig:comp_start_blowup}
\end{figure}

\begin{figure}[tb]
 \centerline{\epsfxsize=2.75in\epsfbox{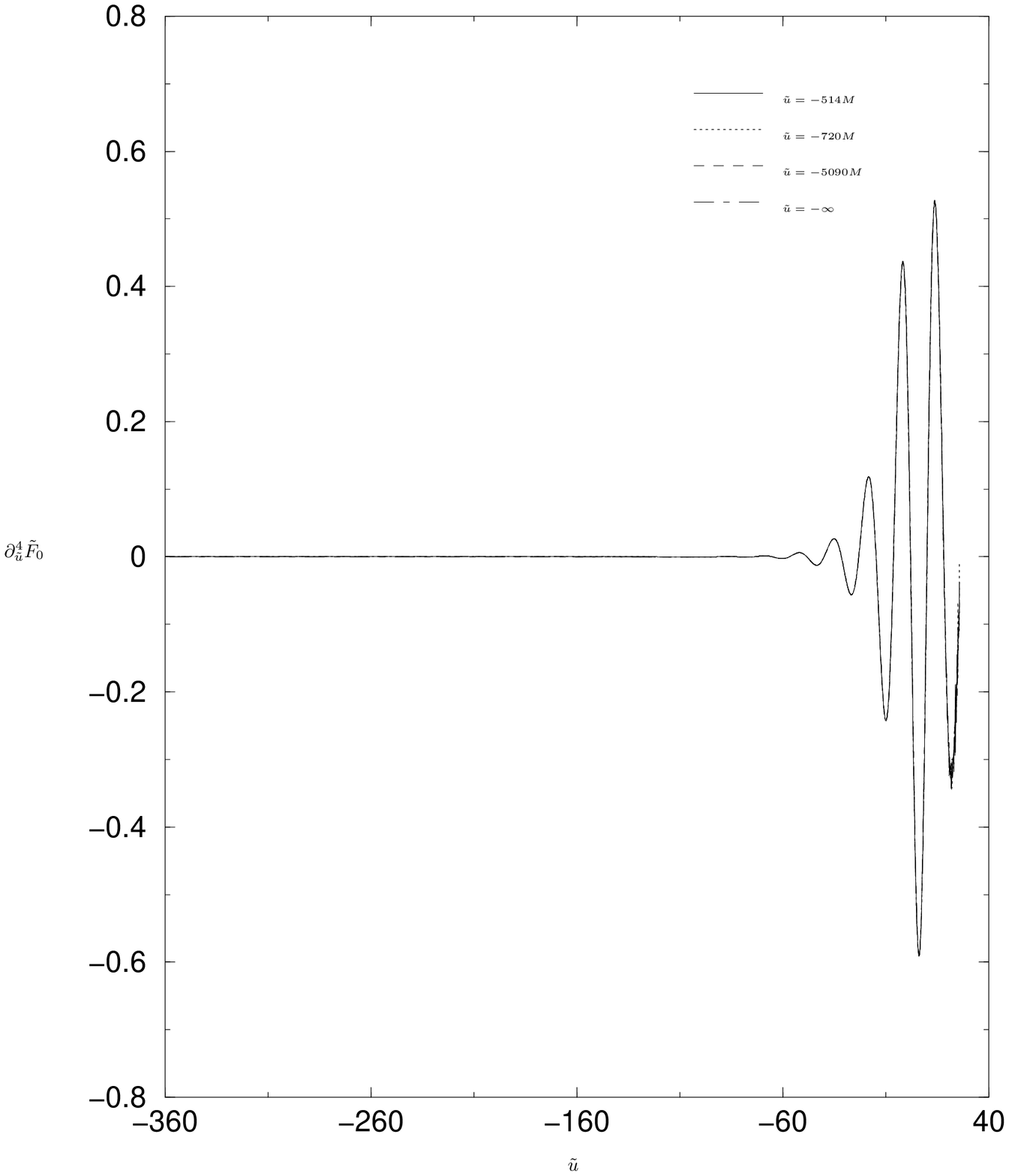}}
  \caption{The 4th $\tilde u$-derivative of Fig~\ref{fig:comp_start_blowup}
           showing the independence on the location of ${\cal J}^-$ .}
 \label{fig:comp_start_4th_deriv}
\end{figure}

Figure~\ref{fig:comp_start_blowup}
shows how the early time polynomial contaminates the genuine
signal at later times, with the effect increasing as ${\cal J}^-$
is located farther in the past. Finally,
Fig.~\ref{fig:comp_start_4th_deriv}
shows that $\partial_{\tilde u}^4 \tilde F_0$ is unaffected by the
artificial polynomial behavior and indicates that a start time of
$\tilde u = -500 M$ is sufficiently early to obtain the correct
Stage II retarded waveform on  $\scri^+$.

Both sets of experiments confirm that the artificial behavior of $\tilde F_0$
at early times has no effect on the accurate calculation of the waveform at
${\cal I}^+$.

\subsection{Late time behavior\label{sec:exp}}

Both the retarded time and advanced time algorithms exhibit exponentially
increasing errors at late times.  This numerical problem arises from the
definition of our evolution variables, which are not designed to deal with late
time features near the future horizon ${\cal H}^+$.  For $r\approx 2M$, our
evolution variables are related to the Israel tetrad variable $F_0$ by
\begin{equation}
       \hat F_0=16M^4 \breve F_0 = \frac {\lambda^2}{4M^2} F_0,
\end{equation}
where the affine parameter $\lambda$ is exponentially related to advanced Bondi
time by
$$\lambda=8 M e^{(\tilde v/4 M) -1} .$$
Because $F_0$ is defined in terms of a
globally well-behaved tetrad, this introduces an exponential
dependence in the evolution variables which influences the numerical
performance in the following way.

Near ${\cal H}^+$, as $r \to 2 M$,  Eqs.~(\ref{eq:teuk_hat_f0})
 and (\ref{eq:teuk_for_brev_f0}) reduce to
\begin{equation}
  \left(\partial_{\tilde v}  -\frac{1}{2 M} \right) \partial_{\tilde u}
      f_0 = 0,
\end{equation}
where $f_0 =\hat F_0 = 16M^4 \breve F_0$. This equation admits outgoing waves
of the form $f_0=f(\tilde u)\exp(\tilde v/(2M))$ in the region inside the
Schwarzschild potential. These exponentially growing waves are forbidden in the
analytic version of our problem by the boundary condition that $\hat F_0=\breve
F_0 =0$ on ${\cal K}^-$ but they are excited by roundoff error. As the
computation proceeds, the interval $\Delta \tilde v$ between  ${\cal K}^-$ and
the peak of the Schwarzschild potential increases and eventually round-off
error in this interval undergoes sufficient exponential magnification to spoil
the exterior evolution. There remains an inner region on ${\cal H}^+$ where the
solution is accurate. For both algorithms, doubling the precision at which the
computation is carried out increases the meaningful computation time. For
example, the double precision runs shown in Figs.~\ref{fig:trunc_deriv} and
\ref{fig:comp_start_4th_deriv} end prematurely due to the onset of numerical
noise but in quadruple precision both runs can be extended through five or
six cycles of quasinormal ringdown.

\section{Close approximation retarded waveforms\label{sec:waveforms}}

\begin{figure}[tb]
 \centerline{\epsfxsize=2.75in\epsfbox{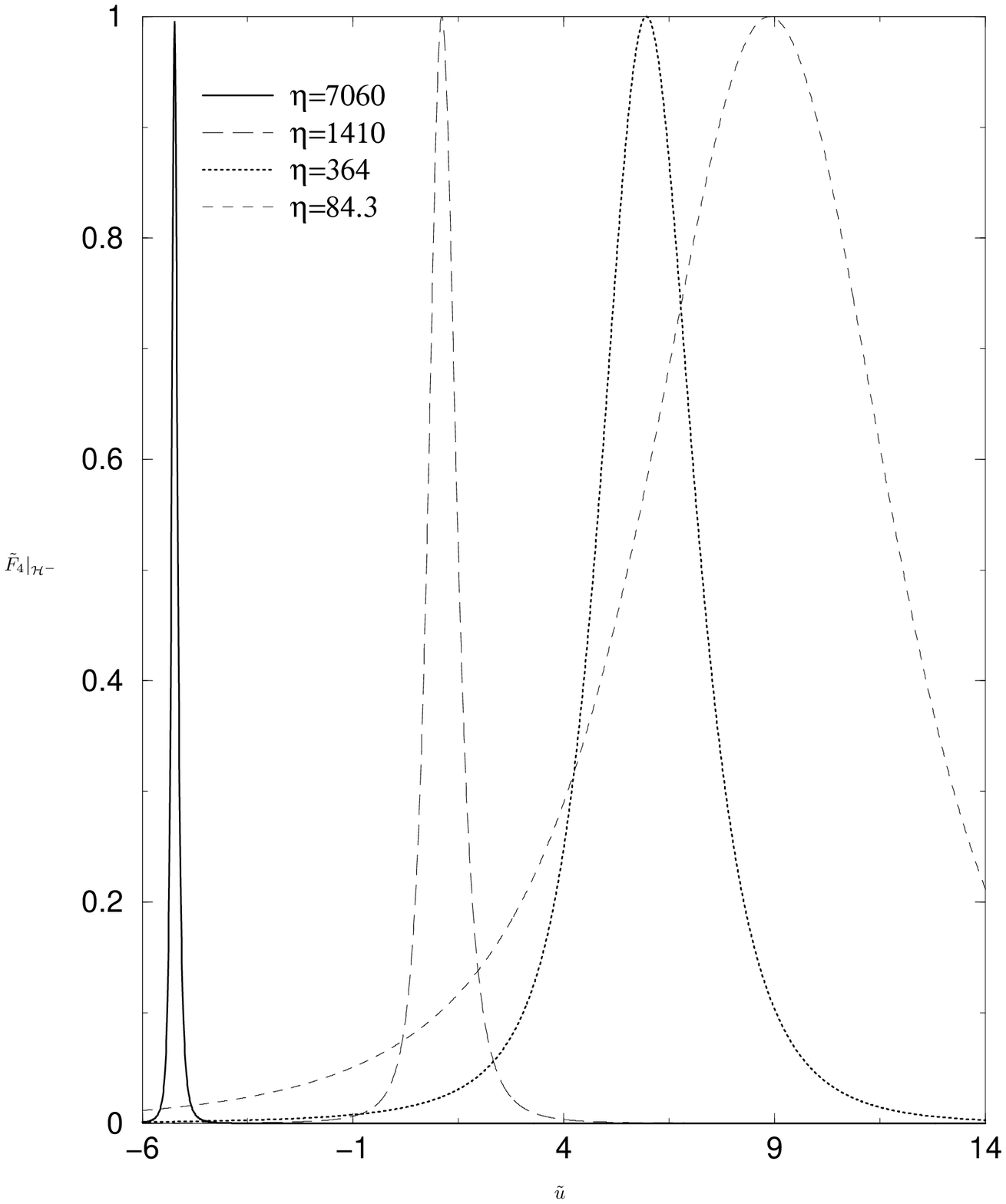}}
  \caption{White hole fission close approximation data
           $\tilde F_4(\tilde u)$ on $\cal H^-$ for
            $\eta=$ 7060, 1410,  364 and 84.3.}
 \label{fig:stage1hor1}
\end{figure}

We begin with a review of the close approximation retarded
waveforms from a white hole fission obtained  in Stage I
\cite{close1}. The close approximation white hole horizon
perturbation ($\tilde F_4|_{{\cal H}^-}$) given by the conformal
model is a pulse with  the following basic properties in terms of
its dependence on the scale invariant parameter $\eta$ (see
Sec.\ref{sec:close}). In the large $\eta$ limit ($\eta > 250$) the
amplitude of the pulse scales quadratically with $\eta$ and the
width  is a monotonically decreasing function of $\eta$ (the width
becomes zero as $\eta \to \infty$). In the small $\eta$ limit
($\eta < 25$) the amplitude scales linearly with $\eta$ but the
shape of the pulse  is independent of $\eta$.
Fig.~\ref{fig:stage1hor1}
shows the horizon data obtained from various large and mid-size
values of $\eta$ (the amplitudes have been  renormalized so that
all pulses have the same height).

\begin{figure}[tb]
 \centerline{\epsfxsize=2.75in\epsfbox{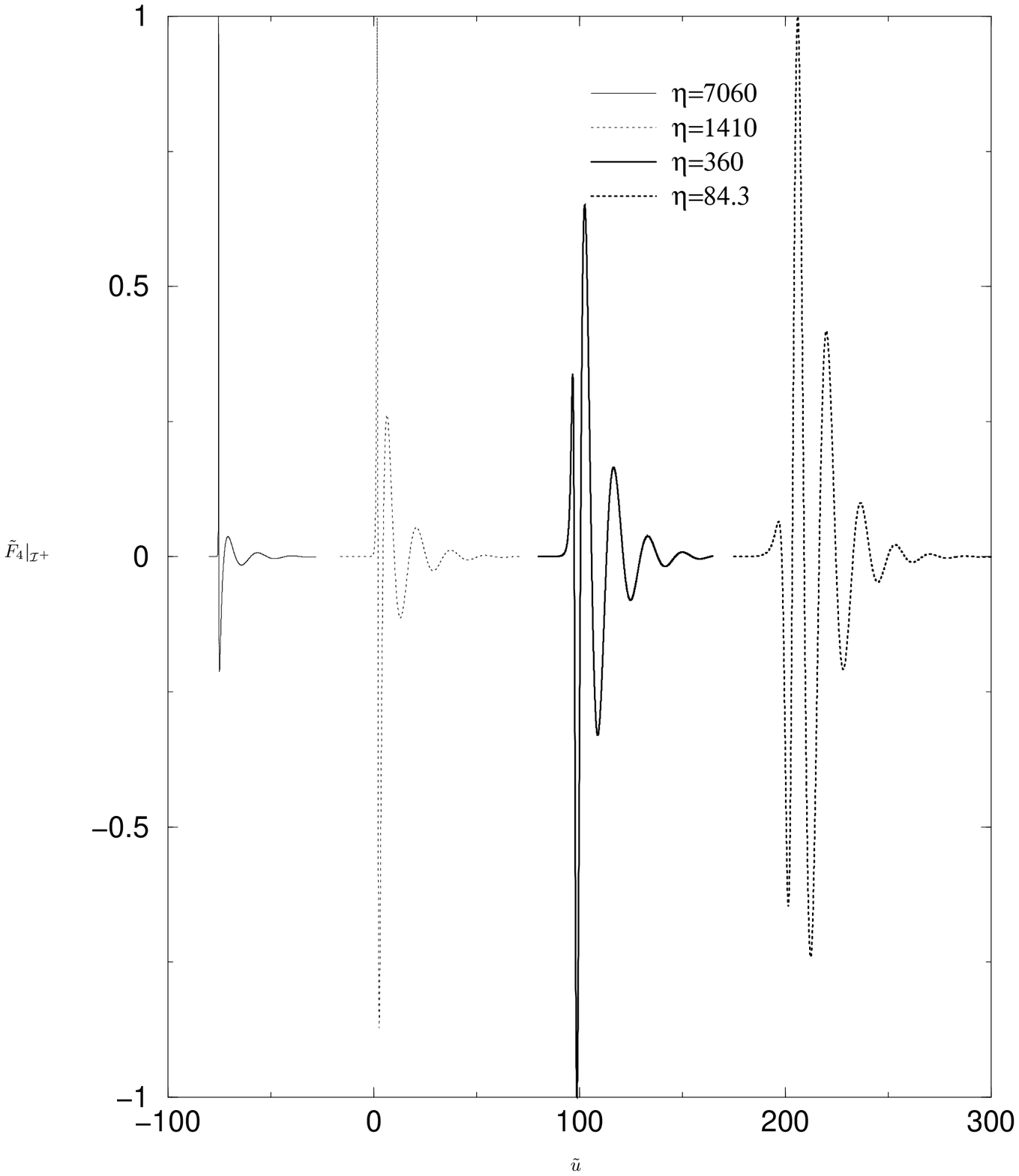}}
  \caption{Close approximation waveforms ${\tilde F_4(\tilde u)}$ on
           $\cal I^+$ for $\eta = 7060$, $1400$, $368$ and $84.3$.}
 \label{fig:stage1wave}
\end{figure}

The associated retarded white hole fission waveforms ($\tilde F_4|_{\scri^+}$)
have a similar dependence on $\eta$. In the large $\eta$ limit the waveform
consists of an extremely sharp initial pulse followed by ringdown and power law
tail. The amplitude of the initial  pulse scales quadratically with $\eta$ and
the width is a monotonically decreasing function of $\eta$. As $\eta$ is
reduced, the initial pulse (signal prior to ringdown) becomes wider and obtains
more structure. In the small $\eta$ limit the  amplitude scales linearly with
$\eta$ and the shape is independent of $\eta$. Fig.~\ref{fig:stage1wave}
shows a  sequence of waveforms from large to mid-size values of
$\eta$.

We now present the results for the retarded solutions from a  black hole fusion
in the close approximation obtained by superposition of the advanced black hole
results (time reversal of the above white hole  solutions) and  the Stage II
results. This superposition yields the retarded waveform and a modified black
hole horizon perturbation. In the following paragraph, we first describe the
black hole horizon perturbations.

\begin{figure}[tb]
 \centerline{\epsfxsize=2.75in\epsfbox{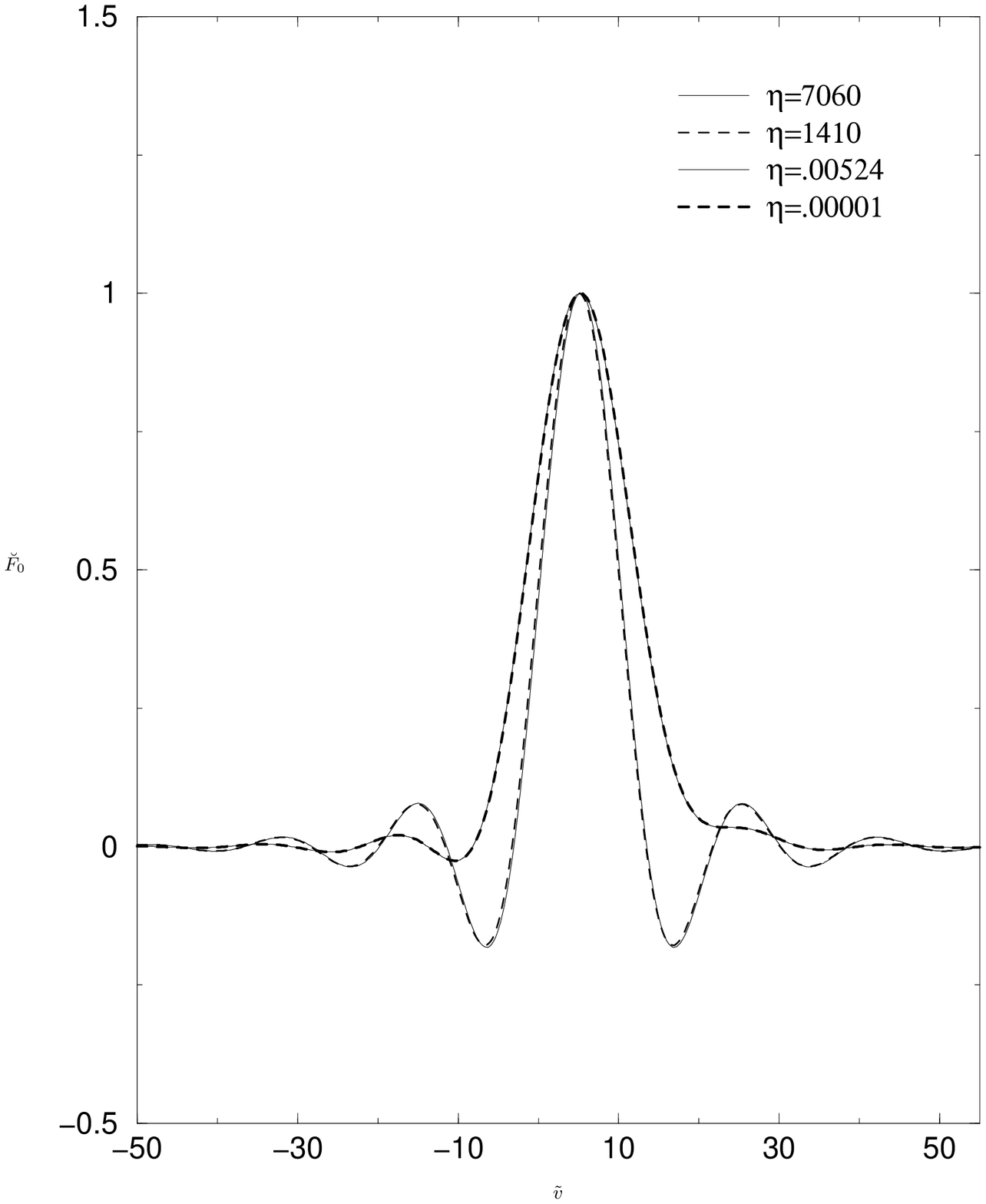}}
  \caption{Final black hole horizon perturbation after superposing stage I
           and II, with amplitudes renormalized so that the peaks match.
           The narrow peak corresponds to $\eta = 7060$ and $\eta = 1410$.}
 \label{fig:conf_hor}
\end{figure}

Figure~\ref{fig:conf_hor}
shows the perturbation on ${\cal H}^+$  for the retarded black
hole solution. Unlike the Stage I perturbation of the white hole
horizon (see Fig.~\ref{fig:stage1hor1}), here the width of the
perturbation is never smaller than a quasinormal period.

The superposition procedure for obtaining the Stage II
perturbation of the black hole horizon turns the very sharp Stage
I pulses associated with large $\eta$ into relatively broad
pulses. The final black hole perturbation approaches a unique
asymptotic shape in both the large and small $\eta$ limits. In
Fig.~\ref{fig:conf_hor}, $\eta = 7060$ and $\eta = 1410$ give the
large $\eta$ shape, while $\eta=.00524$ and $\eta = .00001$ give
the same small $\eta$ shape. Note that the two extremes are
themselves very similar. Consequently all the retarded black hole
waveforms obtained from this procedure are very similar in shape.

One of the most striking features of this new horizon perturbation is the
initial ringup, which is later followed by the expected ringdown due to
scattering off the Schwarzschild potential.  The ringup on $\cal H^+$ results
from the ringup of the data on $\cal J^-$, which in turn results from the
ringdown of the Stage I waveform. This ringup is an unavoidable feature of our
two stage methodology for removing ingoing radiation from the black hole merger.

\begin{figure}[tb]
 \centerline{\epsfxsize=2.75in\epsfbox{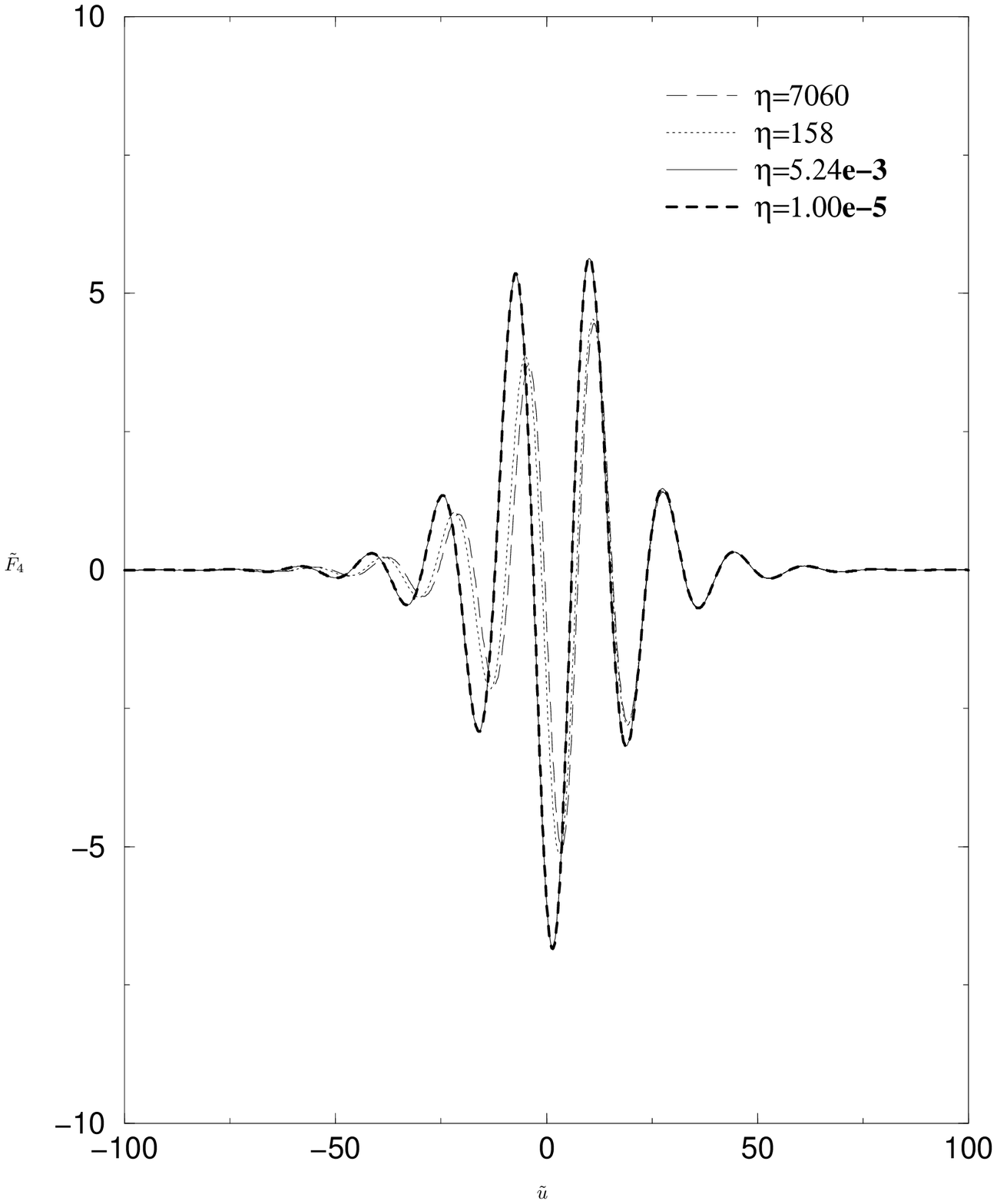}}
 \caption{Black hole retarded waveforms for close approximation data.}
 \label{fig:conf_ret}
\end{figure}

Figure~\ref{fig:conf_ret}
shows the waveforms for close approximation data obtained from the
conformal model. The amplitude of the waveforms scales linearly
with $\eta$ and has been renormalized so that all waveforms show
the same quasinormal decay at late times. Note that, aside from
amplitude, the waveforms have only weak dependence on $\eta$. This
is a great departure from the stage I results that the widths of
the retarded white hole waveforms depend strongly on $\eta$, which
produces extremely narrow pulses for large $\eta$.

It should not be surprising that retarded waveforms from a black
hole merger differ qualitatively from retarded waveforms from a
white hole fission. The fission process is directly observable at
${\cal I}^+$, whereas the merger waveform results indirectly from
the black holes through the preceding collapse of matter or
gravitational wave energy that formed them. This also explains why
the fission waveform should be more sensitive to the parameter
$\eta$ which controls the shape and timescale of the horizon data.
However, the weakness of the dependence of the merger waveform on
$\eta$ is surprising.

\begin{figure}[tb]
 \centerline{\epsfxsize=2.75in\epsfbox{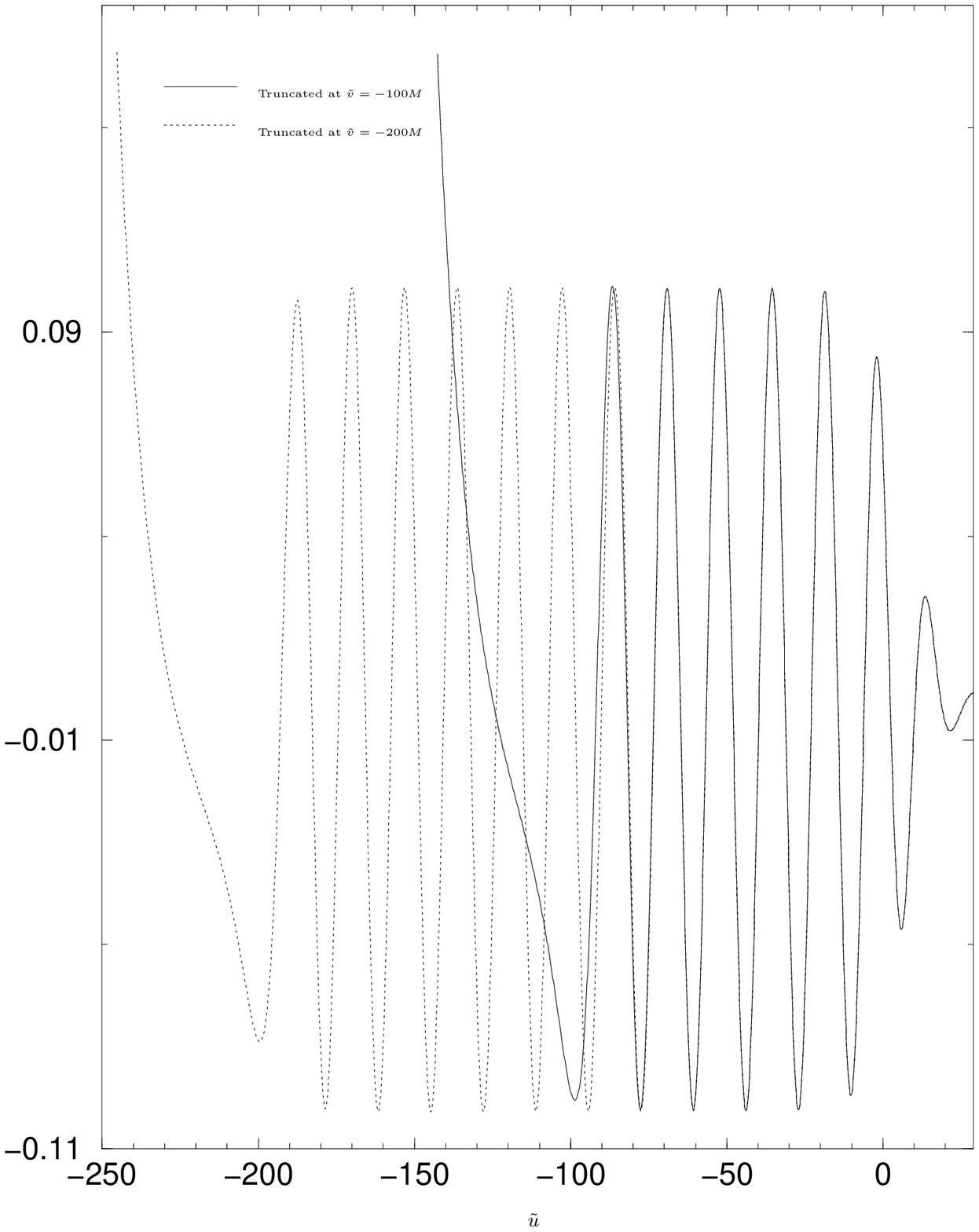}}
  \caption{Ringup of the waveform: $\tilde F_4|_{\scri^+}$ multiplied by
           $\exp -.08896 \tilde u/M$}
 \label{fig:tilde_f_4_ringup_exp}
\end{figure}

The retarded black hole waveforms at ${\cal I}^+$ also  show the
ringup/ringdown pattern. The ringup is due to backscatter of the
ringup on the initial slice ${\cal J}^-$. Interestingly, the
backscattered signal of an $\ell = 2$ quasinormal ringup on $\cal
J^-$ is a ringup on $\scri^+$ with the same $\ell = 2$ time
dependence. Figure~\ref{fig:tilde_f_4_ringup_exp}
shows the ringup phase of the waveform obtained by placing $\cal
K^-$ at $\tilde v = -100M$ and at $\tilde v = -200M$. (The data
for $\tilde v = -200M$ contains more ringup cycles.) In the figure
$\tilde F_4|_{\scri^+}$ is multiplied by $\exp -.08896 \tilde u$
to remove the exponential part of the quasinormal time dependence.
Note that in the  $\tilde v = -200M$ case the ringup begins
exactly $100M$ sooner indicating that it is indeed the ringup of
the initial data that is responsible for the ringup of the
waveform on $\scri^+$.

\section{Discussion}

We have combined the conformal horizon model with a null evolution code to give
a new way to calculate the retarded waveform from a binary black hole merger in
the close approximation. The process of removing ingoing radiation from the
system leads to two major new features in the shape of the close approximation
waveforms for a head-on collision: (i) an initial quasinormal ringup and (ii)
weak sensitivity to the parameter $\eta$ of the model that in some heuristic
sense controls the collision velocity. Feature (ii) has the potential
importance of enabling the design of an efficient template for extracting a
gravitational wave signal from noise.

Similar attempts to remove ingoing radiation from Cauchy evolutions of close
approximation data have not yet been made. Such studies would help clarify
whether the above features are of intrinsic physical origin or an artifact of
our methodology.

The present work is part of an ongoing effort to carry out a similar two stage
characteristic evolution to provide the waveforms from coalescing black holes
in the nonlinear regime. Because this is an unexplored area of binary black
hole physics, this perturbative study of the head-on collision provides a
preliminary physical check which will be useful in extending the work to the
nonlinear and nonaxisymmetric case, where inspiraling black holes can be
treated.

\begin{acknowledgments}
We thank Manuela Campanelli for numerous helpful discussions. This work has been
supported by NSF grants PHY 9800731 and PHY 9988663 to the University of
Pittsburgh. Y.~Z. thanks the Albert-Einstein-Institut for hospitality. Computer
time for this project has been provided by the Pittsburgh Supercomputing Center.
\end{acknowledgments}


\begin{thebibliography}{10}

\bibitem{close1}
M. Campanelli, R. G\'omez, S.Husa, J. Winicour and
Y. Zlochower, Phys. Rev. D  {\bf 63} 124013 (2001).

\bibitem{pp}
R, H. Price and J. Pullin, Phys. Rev. Lett. {\bf 72}, 3297 (1994).

\bibitem{price97}
Z. Andrade and R, Price, Phys. Rev. D {\bf 56}, 6336 (1997).

\bibitem{Campanelli98b}
M. Campanelli, W. Krivan, and C.~O. Lousto, Phys. Rev. D
{\bf 58}, 024016 (1998).

\bibitem{Campanelli98a}
M. Campanelli and C.~O. Lousto, Phys. Rev. D {\bf 58}, 024015
(1998).

\bibitem{Campanelli98c}
M. Campanelli, C.~O. Lousto, J. Baker, G. Khanna, and
J. Pullin, Phys. Rev. D {\bf 58}, 084019 (1998).

\bibitem{price99}
R. Glaser, O. Nicasio, R, Price and J. Pullin, Phys. Rev. D {\bf 59}, 044024
(1999).

\bibitem{price99l}
W. Krivan and R, Price, Phys. Rev. Lett. {\bf 82}, 1358 (1999).

\bibitem{price99l2}
G. Khanna, J. Baker, R. Glaser, P. Laguna, O. Nicasio, H.-P. Nollert.
R, Price and J. Pullin, Phys. Rev. Lett. {\bf 83}, 3581 (1999).

\bibitem{Campanelli99}
M. Campanelli and C.~O. Lousto, Phys. Rev. D {\bf 59}, 124022
(1999).

\bibitem{Baker99a}
J. Baker, S. Brandt, M. Campanelli, C.~O. Lousto, E.
Seidel, and R. Takahashi, Phys. Rev. D {\bf 62}, 127701 (2000).

\bibitem{Baker2000b}
J. Baker, B. Br\"ugmann, M. Campanelli, and C.~O. Lousto,
Class. Quantum Grav. {\bf 17}, L149 (2000).

\bibitem{kyoto}
J. Winicour, Prog. Theor. Phys. Supp. {\bf 136}, 57 (1999).

\bibitem{compnull}
R. G\'omez, S. Husa, and J. Winicour,  Phys. Rev. D
{\bf 64} (2001) 024010.

\bibitem{fiss}
R. G\'omez, S. Husa, L. Lehner, and J. Winicour,
{\it Gravitational Waves from a Fissioning White Hole}, in preparation.

\bibitem{high}
N.~T. Bishop, R. G\'omez, L. Lehner, M. Maharaj, J. Winicour,
Phys. Rev. D {\bf 56} 6298 (1997).

\bibitem{reduced}
R. G\'omez, Phys. Rev. D {\bf 64} 024007, (2001).

\bibitem{sachsdn}
R. Sachs, J. Math. Phys. {\bf 3}, 908 (1962).

\bibitem{haywdn}
S.~A. Hayward, Class. Quantum Grav. {\bf 10}, 779 (19993).

\bibitem{fried81}
H. Friedrich, Proc. R. Soc. London A {\bf 378}, 401 (1981).

\bibitem{rend}
A.~H. Rendall, Proc. R. Soc. London A {\bf 427}, 221 (1990).

\bibitem{Teukolsky73}
S.~A. Teukolsky, Astrophys. J. {\bf 185}, 635 (1973).

\bibitem{israel}
W. Israel, Phys. Rev. {\bf 143}, 1016 (1966).

\bibitem{Newman62a}
E.~T. Newman and R. Penrose, J. Math. Phys. {\bf 3}, 566 (1962).

\bibitem{ndata}
L. Lehner, N.~T. Bishop, R. G\'omez, B. Szil\'agyi, and
J. Winicour, Phys. Rev. D {\bf 60}, 044005 (1999).

\bibitem{asym}
S. Husa and J. Winicour, Phys. Rev. D {\bf 60}, 084019 (1999).

\bibitem{nu}
E.~T.  Newman and T.~W.~J. Unti, J. Math. Phys. {\bf 5}, 891 (1962).

\bibitem{bondi}
H. Bondi, M.~J.~G. van der Burg and A.~W.~K. Metzner,
Proc. R. Soc. London A {\bf 269} 21, (1962).

\bibitem{sachs}
R.K. Sachs, Proc. R. Soc. London A {\bf 270} 103, 1962.

\bibitem{cbf}
L. Tamburino and J. Winicour, Phys. Rev. {\bf 150}, 1039 (1966).

\bibitem{berndtub}
B.~G. Schmidt, talk at the T\"ubingen workshop (April, 2001).


\end{thebibliography}

\end{document}